\pdfoutput=1
\documentclass[12pt,a4paper]{article}

\usepackage{ifthen} 
\usepackage{multirow}
\newboolean{pdflatex}
\setboolean{pdflatex}{true} 

\newboolean{articletitles}
\setboolean{articletitles}{true} 

\newboolean{uprightparticles}
\setboolean{uprightparticles}{false} 

\def\paperauthors{LHCb collaboration} 
\def\paperasciititle{BuLcLcK} 
\def\papertitle{Study of the $\BtoLcLcK$ decay} 
\def\paperkeywords{{High Energy Physics}, {LHCb}} 
\def\papercopyright{\the\year\ CERN for the benefit of the LHCb collaboration} 
\def\paperlicence{CC BY 4.0 licence}
\def\paperlicenceurl{https://creativecommons.org/licenses/by/4.0/}

\usepackage[top=1in, bottom=1.25in, left=1in, right=1in]{geometry}

\usepackage{microtype}
\usepackage{lineno}  
\usepackage{xspace} 
\usepackage{caption} 

\usepackage{graphicx}  
\usepackage{color}
\usepackage{colortbl}
\graphicspath{{./figs/}} 
\usepackage{epsfig}

\usepackage{amsmath} 
\usepackage{amssymb}
\usepackage{amsfonts}
\usepackage{upgreek} 
\usepackage{cancel}

\newcommand*\patchAmsMathEnvironmentForLineno[1]{%
\expandafter\let\csname old#1\expandafter\endcsname\csname #1\endcsname
\expandafter\let\csname oldend#1\expandafter\endcsname\csname
end#1\endcsname
 \renewenvironment{#1}%
   {\linenomath\csname old#1\endcsname}%
   {\csname oldend#1\endcsname\endlinenomath}%
}
\newcommand*\patchBothAmsMathEnvironmentsForLineno[1]{%
  \patchAmsMathEnvironmentForLineno{#1}%
  \patchAmsMathEnvironmentForLineno{#1*}%
}
\AtBeginDocument{%
\patchBothAmsMathEnvironmentsForLineno{equation}%
\patchBothAmsMathEnvironmentsForLineno{align}%
\patchBothAmsMathEnvironmentsForLineno{flalign}%
\patchBothAmsMathEnvironmentsForLineno{alignat}%
\patchBothAmsMathEnvironmentsForLineno{gather}%
\patchBothAmsMathEnvironmentsForLineno{multline}%
\patchBothAmsMathEnvironmentsForLineno{eqnarray}%
}

\usepackage{hyperxmp}

\usepackage[pdftex,
            pdfauthor={\paperauthors},
            pdftitle={\paperasciititle},
            pdfkeywords={\paperkeywords},
            pdfcopyright={Copyright (C) \papercopyright},
            pdflicenseurl={\paperlicenceurl}]{hyperref}

\usepackage[colorinlistoftodos,textsize=scriptsize]{todonotes}

\usepackage[bottom,flushmargin,hang,multiple]{footmisc}

\usepackage[all]{hypcap} 

\usepackage{xspace} 
\usepackage{upgreek}

\def\lhcb   {\mbox{LHCb}\xspace}

\def\babar  {\mbox{BaBar}\xspace}
\def\belle  {\mbox{Belle}\xspace}

\def\MagUp {\mbox{\em Mag\kern -0.05em Up}\xspace}

\ifthenelse{\boolean{uprightparticles}}%
{

 \def\Pmu         {\ensuremath{\upmu}\xspace}

 \def\Ppi         {\ensuremath{\uppi}\xspace}

 \def\Ppsi        {\ensuremath{\uppsi}\xspace}

 \def\PDelta      {\ensuremath{\Delta}\xspace}                 
 \def\PXi         {\ensuremath{\Xi}\xspace}                 
 \def\PLambda     {\ensuremath{\Lambda}\xspace}                 
 \def\PSigma      {\ensuremath{\Sigma}\xspace}                 
 \def\POmega      {\ensuremath{\Omega}\xspace}                 
 \def\PUpsilon    {\ensuremath{\Upsilon}\xspace}
 \let\oldPi\Pi
 \def\PPi         {\ensuremath{\oldPi}\xspace}

 \def\PB      {\ensuremath{\mathrm{B}}\xspace}                 
                  
 \def\PD      {\ensuremath{\mathrm{D}}\xspace}

 \def\PJ      {\ensuremath{\mathrm{J}}\xspace}                 
 \def\PK      {\ensuremath{\mathrm{K}}\xspace}

 \def\Pb      {\ensuremath{\mathrm{b}}\xspace}                 
 \def\Pc      {\ensuremath{\mathrm{c}}\xspace}                 
                  
 \def\Pe      {\ensuremath{\mathrm{e}}\xspace}

 \def\Pi      {\ensuremath{\mathrm{i}}\xspace}

 \def\Pp      {\ensuremath{\mathrm{p}}\xspace}

 \def\Ps      {\ensuremath{\mathrm{s}}\xspace}

 \def\thebaroffset{0.0em}
}
{

 \def\Pmu         {\ensuremath{\mu}\xspace}

 \def\Ppi         {\ensuremath{\pi}\xspace}

 \def\Ppsi        {\ensuremath{\psi}\xspace}                 
                  
 \mathchardef\PDelta="7101
 \mathchardef\PXi="7104
 \mathchardef\PLambda="7103
 \mathchardef\PSigma="7106
 \mathchardef\POmega="710A
 \mathchardef\PUpsilon="7107
 \mathchardef\PPi="7105
                  
 \def\PB      {\ensuremath{B}\xspace}                 
                  
 \def\PD      {\ensuremath{D}\xspace}

 \def\PJ      {\ensuremath{J}\xspace}                 
 \def\PK      {\ensuremath{K}\xspace}

 \def\Pb      {\ensuremath{b}\xspace}                 
 \def\Pc      {\ensuremath{c}\xspace}                 
                  
 \def\Pe      {\ensuremath{e}\xspace}

 \def\Pi      {\ensuremath{i}\xspace}

 \def\Pp      {\ensuremath{p}\xspace}

 \def\Ps      {\ensuremath{s}\xspace}

 \def\thebaroffset{0.18em}
}
\newcommand{\offsetoverline}[2][\thebaroffset]{\kern #1\overline{\kern -#1 #2}}%

\makeatletter
\ifcase \@ptsize \relax
  \newcommand{\miniscule}{\@setfontsize\miniscule{4}{5}}
\or
  \newcommand{\miniscule}{\@setfontsize\miniscule{5}{6}}
\or
  \newcommand{\miniscule}{\@setfontsize\miniscule{5}{6}}
\fi
\makeatother

\DeclareRobustCommand{\optbar}[1]{\shortstack{{\miniscule (\rule[.5ex]{1.25em}{.18mm})}
  \\ [-.7ex] $#1$}}

\def\en         {{\ensuremath{\Pe^-}}\xspace}   
\def\ep         {{\ensuremath{\Pe^+}}\xspace}

\def\mumu       {{\ensuremath{\Pmu^+\Pmu^-}}\xspace}

\def\squark    {{\ensuremath{\Ps}}\xspace}

\def\cquark    {{\ensuremath{\Pc}}\xspace}

\def\bquark    {{\ensuremath{\Pb}}\xspace}

\def\pion   {{\ensuremath{\Ppi}}\xspace}
\def\piz    {{\ensuremath{\pion^0}}\xspace}
\def\pip    {{\ensuremath{\pion^+}}\xspace}
\def\pim    {{\ensuremath{\pion^-}}\xspace}

\def\kaon    {{\ensuremath{\PK}}\xspace}

\def\KorKbar {\kern \thebaroffset\optbar{\kern -\thebaroffset \PK}{}\xspace}

\def\Kp      {{\ensuremath{\kaon^+}}\xspace}
\def\Km      {{\ensuremath{\kaon^-}}\xspace}

\def\KS      {{\ensuremath{\kaon^0_{\mathrm{S}}}}\xspace}

\def\D       {{\ensuremath{\PD}}\xspace}

\def\DorDbar {\kern \thebaroffset\optbar{\kern -\thebaroffset \PD}\xspace}

\def\Dp      {{\ensuremath{\D^+}}\xspace}
\def\Dm      {{\ensuremath{\D^-}}\xspace}

\def\DpDm    {\ensuremath{\Dp {\kern -0.16em \Dm}}\xspace}

\def\B       {{\ensuremath{\PB}}\xspace}

\def\BorBbar {\kern \thebaroffset\optbar{\kern -\thebaroffset \PB}\xspace}

\def\Bd      {{\ensuremath{\B^0}}\xspace}

\def\BdorBdbar {\kern \thebaroffset\optbar{\kern -\thebaroffset \Bd}\xspace}

\def\Bub     {{\ensuremath{\B^-}}\xspace}

\def\Bm      {{\ensuremath{\Bub}}\xspace}

\def\Bs      {{\ensuremath{\B^0_\squark}}\xspace}

\def\BsorBsbar {\kern \thebaroffset\optbar{\kern -\thebaroffset \Bs}\xspace}

\def\jpsi     {{\ensuremath{{\PJ\mskip -3mu/\mskip -2mu\Ppsi}}}\xspace}

\def\Y#1S{\ensuremath{\PUpsilon{(#1S)}}\xspace}

\def\proton      {{\ensuremath{\Pp}}\xspace}
\def\antiproton  {{\ensuremath{\overline \proton}}\xspace}

\def\Lz          {{\ensuremath{\PLambda}}\xspace}
\def\Lbar        {{\ensuremath{\offsetoverline{\PLambda}}}\xspace}
\def\LorLbar     {\kern \thebaroffset\optbar{\kern -\thebaroffset \PLambda}\xspace}

\def\Sigmares    {{\ensuremath{\PSigma}}\xspace}

\def\Xires       {{\ensuremath{\PXi}}\xspace}

\def\Omegares    {{\ensuremath{\POmega}}\xspace}

\def\Lc          {{\ensuremath{\Lz^+_\cquark}}\xspace}
\def\Lcbar       {{\ensuremath{\Lbar{}^-_\cquark}}\xspace}
\def\Xic         {{\ensuremath{\Xires_\cquark}}\xspace}
\def\Xicz        {{\ensuremath{\Xires^0_\cquark}}\xspace}

\def\Omegac      {{\ensuremath{\Omegares^0_\cquark}}\xspace}

\def\Lb           {{\ensuremath{\Lz^0_\bquark}}\xspace}

\def\BF         {{\ensuremath{\mathcal{B}}}\xspace}

\newcommand{\decay}[2]{\ensuremath{#1\!\to #2}\xspace} 

\def\to                 {\ensuremath{\rightarrow}\xspace}

\newcommand{\etot}{{\ensuremath{\varepsilon_{\mathrm{ tot}}}}\xspace}

\def\AT#1     {\ensuremath{A_{\mathrm{T}}^{#1}}\xspace}           

\def\C#1      {\ensuremath{\mathcal{C}_{#1}}\xspace}                       
\def\Cp#1     {\ensuremath{\mathcal{C}_{#1}^{'}}\xspace}                    
\def\Ceff#1   {\ensuremath{\mathcal{C}_{#1}^{\mathrm{(eff)}}}\xspace}        
\def\Cpeff#1  {\ensuremath{\mathcal{C}_{#1}^{'\mathrm{(eff)}}}\xspace}       
\def\Ope#1    {\ensuremath{\mathcal{O}_{#1}}\xspace}                       
\def\Opep#1   {\ensuremath{\mathcal{O}_{#1}^{'}}\xspace}                    

\newcommand{\nospaceunit}[1]{\ensuremath{\text{#1}}}       
\newcommand{\aunit}[1]{\ensuremath{\text{\,#1}}}       

\newcommand{\tev}{\aunit{Te\kern -0.1em V}\xspace}
\newcommand{\gev}{\aunit{Ge\kern -0.1em V}\xspace}
\newcommand{\mev}{\aunit{Me\kern -0.1em V}\xspace}
\newcommand{\kev}{\aunit{ke\kern -0.1em V}\xspace}
\newcommand{\ev}{\aunit{e\kern -0.1em V}\xspace}
 
\newcommand{\mevc}{\ensuremath{\aunit{Me\kern -0.1em V\!/}c}\xspace}
\newcommand{\gevc}{\ensuremath{\aunit{Ge\kern -0.1em V\!/}c}\xspace}
\newcommand{\mevcc}{\ensuremath{\aunit{Me\kern -0.1em V\!/}c^2}\xspace}
\newcommand{\gevcc}{\ensuremath{\aunit{Ge\kern -0.1em V\!/}c^2}\xspace}

\def\mum  {\ensuremath{\,\upmu\nospaceunit{m}}\xspace}

\def\fb   {\ensuremath{\aunit{fb}}\xspace}
\def\invfb   {\ensuremath{\fb^{-1}}\xspace}

\newcommand{\chisq}{\ensuremath{\chi^2}\xspace}

\newcommand{\chisqip}{\ensuremath{\chi^2_{\text{IP}}}\xspace}

\def\gsim{{~\raise.15em\hbox{$>$}\kern-.85em
          \lower.35em\hbox{$\sim$}~}\xspace}
\def\lsim{{~\raise.15em\hbox{$<$}\kern-.85em
          \lower.35em\hbox{$\sim$}~}\xspace}

\def\pt         {\ensuremath{p_{\mathrm{T}}}\xspace}

\def\ptot       {\ensuremath{p}\xspace}

\def\evtgen     {\mbox{\textsc{EvtGen}}\xspace}

\def\geant      {\mbox{\textsc{Geant4}}\xspace}

\def\photos     {\mbox{\textsc{Photos}}\xspace}

\def\pythia     {\mbox{\textsc{Pythia}}\xspace}

\def\tell1  {TELL1\xspace}
\def\ukl1   {UKL1\xspace}

\newcommand{\eg}{\mbox{\itshape e.g.}\xspace}

\newcommand{\vs}{\mbox{\itshape vs.}\xspace}

\newcommand{\lhcborcid}[1]{\href{https://orcid.org/#1}{\hspace*{0.1em}\raisebox{-0.45ex}{\includegraphics[width=1em]{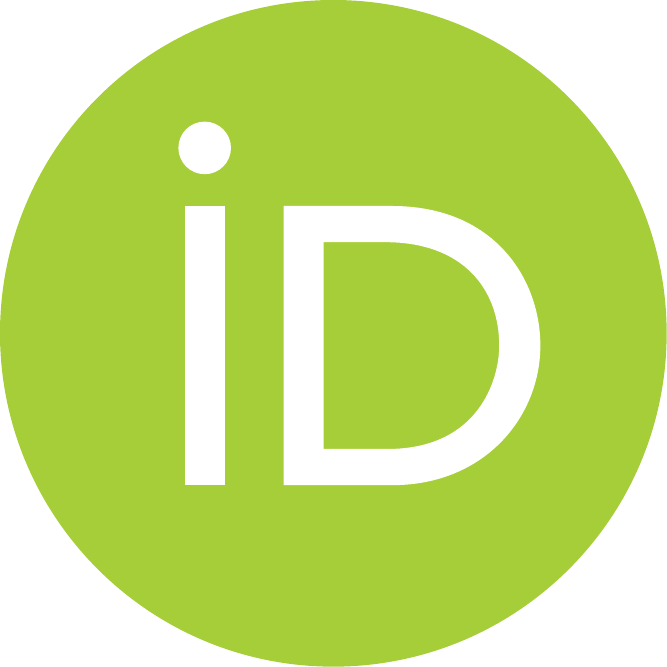}}}}

\usepackage{cite} 
\usepackage{mciteplus}

\newcommand{\BtoLcLcK}{\Bm \to \Lc \Lcbar \Km}
\newcommand{\BdtoLcLcKS}{B^0 \to \Lc \Lcbar \KS}
\newcommand{\BtoDDK}{\Bm \to \Dp \Dm \Km}

\newcommand{\LcLc}{\Lc\Lcbar}

\newcommand{\nevt}{N_{\mathrm{evt}}}
\newcommand{\phsp}{\mathrm{phsp}}
\newcommand{\bkg}{\mathrm{bkg}}

\usepackage{longtable} 
\begin{document}

\renewcommand{\thefootnote}{\fnsymbol{footnote}}
\setcounter{footnote}{1}


\begin{titlepage}
\pagenumbering{roman}

\vspace*{-1.5cm}
\centerline{\large EUROPEAN ORGANIZATION FOR NUCLEAR RESEARCH (CERN)}
\vspace*{1.5cm}
\noindent
\begin{tabular*}{\linewidth}{lc@{\extracolsep{\fill}}r@{\extracolsep{0pt}}}
\ifthenelse{\boolean{pdflatex}}
{\vspace*{-1.5cm}\mbox{\!\!\!\includegraphics[width=.14\textwidth]{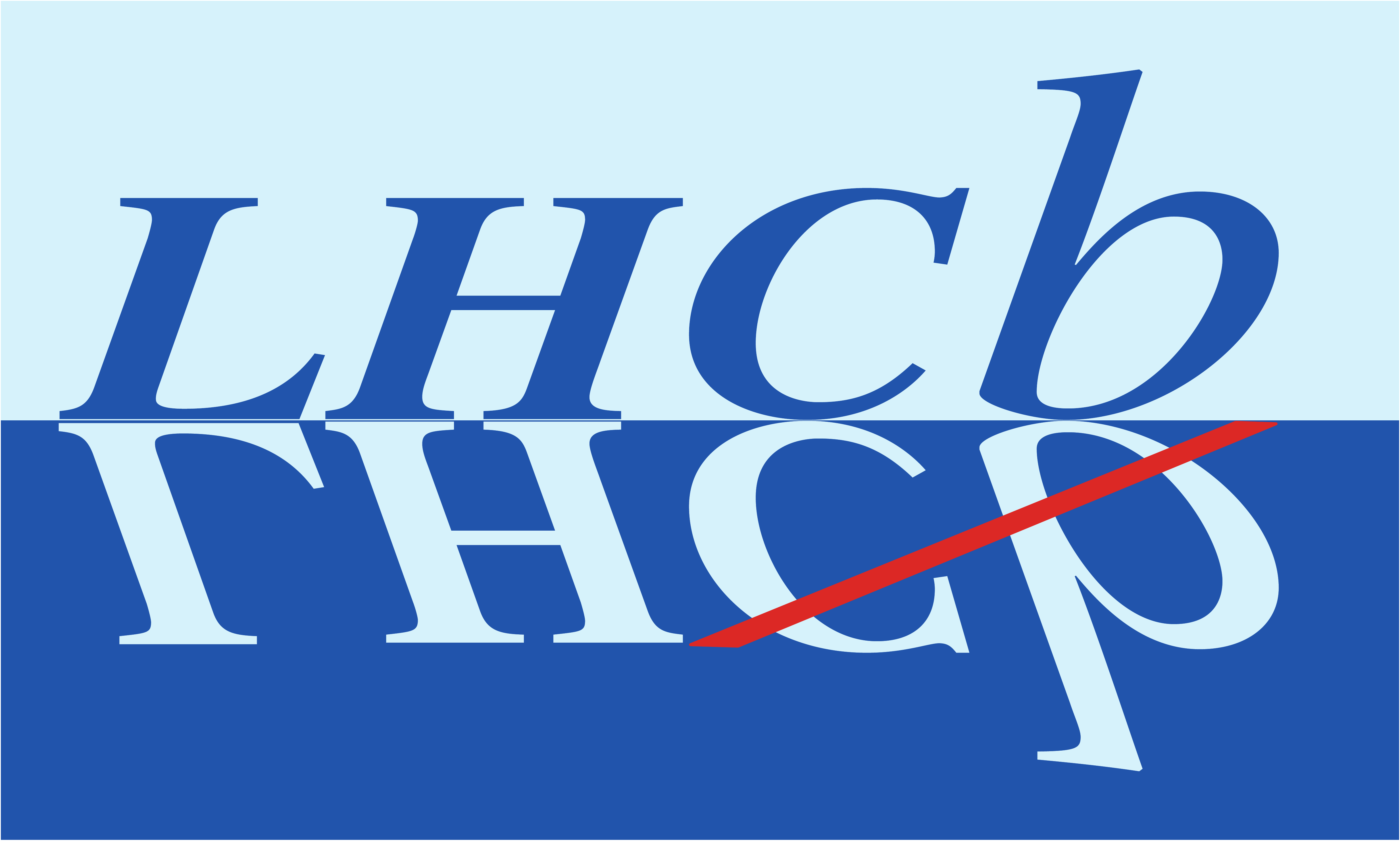}} & &}%
{\vspace*{-1.2cm}\mbox{\!\!\!\includegraphics[width=.12\textwidth]{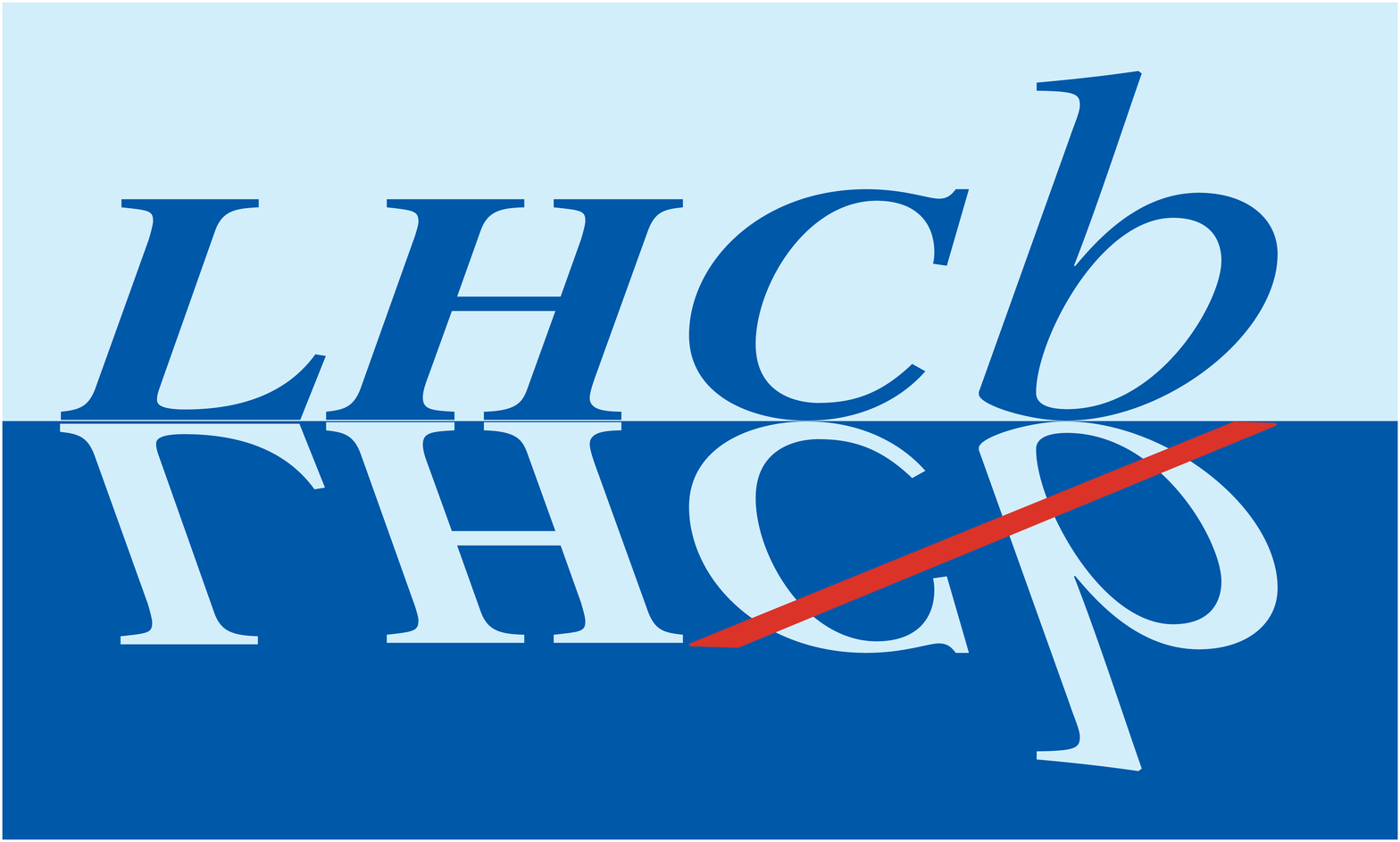}} & &}%
\\
 & & CERN-EP-2022-196 \\  
 & & LHCb-PAPER-2022-028 \\  
 & & 12 Aug 2023 \\ 
 & & \\
\end{tabular*}

\vspace*{4.0cm}

{\normalfont\bfseries\boldmath\huge
\begin{center}
  \papertitle 
\end{center}
}

\vspace*{0.5cm}

\begin{center}
\paperauthors \footnote{Authors are listed at the end of this paper.}
\end{center}

\vspace*{0.5cm}

\begin{abstract}
  \noindent
  The decay $\BtoLcLcK$ is studied in proton-proton collisions
  at a center-of-mass energy of $\sqrt{s}=13$ TeV using data corresponding to 
  an integrated luminosity of \mbox{5 $\invfb$} collected by the LHCb experiment.
  In the $\Lc \Km$ system, the $\Xic(2930)^{0}$ state observed at the
  \babar and \belle experiments is resolved into two narrower states, 
  $\Xic(2923)^{0}$ and $\Xic(2939)^{0}$,
  whose masses and widths are measured to be
  \begin{align*}
    m(\Xic(2923)^{0}) & = 2924.5 \pm 0.4 \pm 1.1 \mev,  \\
    m(\Xic(2939)^{0}) & =  2938.5 \pm 0.9 \pm 2.3 \mev, \\
    \Gamma(\Xic(2923)^{0}) & = \phantom{000}4.8 \pm 0.9 \pm 1.5 \mev,\\
    \Gamma(\Xic(2939)^{0}) & = \phantom{00}11.0 \pm 1.9 \pm 7.5 \mev, 
  \end{align*}
  where the first uncertainties are statistical and the second systematic.
  The results are consistent with a previous LHCb measurement using a prompt $\Lc \Km$ sample.
  Evidence of a new $\Xic(2880)^{0}$ state is found with a local significance of
  $3.8\,\sigma$, whose mass and width are measured to be
  \mbox{$2881.8 \pm 3.1 \pm 8.5\mev$} and \mbox{$12.4 \pm 5.3 \pm 5.8 \mev$}, respectively.
  In addition, evidence of a new decay mode $\Xic(2790)^{0} \to \Lc \Km$ is found
  with a significance of $3.7\,\sigma$.
  The relative branching fraction of $\BtoLcLcK$ with respect to the $\BtoDDK$ decay 
  is measured to be \mbox{$2.36 \pm 0.11 \pm 0.22 \pm 0.25$}, where the first uncertainty is statistical,
  the second systematic and the third originates from the branching fractions of charm hadron decays. 
\end{abstract}

\vspace*{0.5cm}

\begin{center}
  Published in
  Phys.~Rev.~D 108, (2023) 012020
\end{center}

\vspace{\fill}

{\footnotesize 
\centerline{\copyright~\papercopyright. \href{\paperlicenceurl}{\paperlicence}.}}
\vspace*{2mm}

\end{titlepage}


\newpage
\setcounter{page}{2}
\mbox{~}
%
%
%
%


\renewcommand{\thefootnote}{\arabic{footnote}}
\setcounter{footnote}{0}

\cleardoublepage


\pagestyle{plain} 
\setcounter{page}{1}
\pagenumbering{arabic}


\section{Introduction}
\label{sec:introduction}
The quark model predicts a rich spectrum of singly charmed baryons.
However, many states, especially those containing at least one $s$ quark
(\eg $\Xicz$, $\Omegac$),
have not been experimentally established.
An effective approach to search for charm baryons is in $\bquark$-hadron decays. 
Upon the observation of the $\BtoLcLcK$ decay,\footnote{The inclusion of charge-conjugate processes is implied throughout the paper.}
the \babar collaboration reported evidence of a $\Xic(2930)^{0}$ state
in the $\Lc \Km$ system with a low significance\cite{Aubert:2007eb};
and later the \belle collaboration confirmed this with a larger yield\cite{Li:2017uvv}. 
Recently, the \belle collaboration also found evidence of its charged counterpart in the $\BdtoLcLcKS$
decay\cite{Li:2018fmq}.
The LHCb experiment, with a much higher $B$ meson yield and excellent momentum resolution,
should be able to confirm the $\Xic(2930)^{0}$ state by examining the $\BtoLcLcK$ decay, 
precisely measure its properties and resolve a potential finer structure.
   
Another way to look for charmed baryons is to search for their production in primary $pp$ interactions, recently exploited by the LHCb experiment~\cite{LHCb-PAPER-2017-002,LHCb-PAPER-2017-018,LHCb-PAPER-2018-026,LHCb-PAPER-2020-004}.
A search for excited $\Xicz$ baryons in prompt $\Lc\Km$ pairs performed at the  LHCb experiment
found that there are two narrow states, $\Xic(2923)^{0}$ and $\Xic(2939)^{0}$, at the position
of $\Xic(2930)^{0}$ observed at $B$ factories, with the presence of an additional $\Xic(2965)^{0}$ structure at higher mass\cite{LHCb-PAPER-2020-004}.
At lower mass, a peaking structure is seen at around 2880\mev,\footnote{Natural units with $\hbar=c=1$ are used throughout this paper.} noted as $\Xic(2880)^{0}$,
which is obscured by partially reconstructed background but is in the position of an expected resonance~\cite{Wang:2020gkn}.
The complexity of the current experimental situation motivates the study of the $\BtoLcLcK$ decay at the LHCb experiment.
Although the exclusive reactions have lower yield compared to the prompt search, the background level is expected to be lower.

Furthermore, the $\BtoLcLcK$ decay provides a unique opportunity to search for possible exotic candidates in
the $\LcLc$ and $\Lcbar \Km$ spectra. The \belle collaboration observed a significant near-threshold enhancement in the $\Lc\Lcbar$ mass spectrum through the reaction
$\ep\en \to \Lc\Lcbar\gamma_{\mathrm{ISR}}$, where $\gamma_{\mathrm{ISR}}$ indicates a high momentum photon emitted in the initial state radiation process~\cite{Pakhlova:2008vn}.
The BESIII experiment recently measured the $\ep\en \to \Lc\Lcbar$ cross-section with high precision,
and an enhanced value near threshold was reported~\cite{Ablikim:2017lct}.
If the enhancement of the $\Lc\Lcbar$ production around $4600\mev$ is due to the presence of a resonance,
it might be confirmed in a decay of the type $\BtoLcLcK$, despite the different production mechanism.
The $\Lcbar \Km$ system does not form resonances in the conventional quark model.

In this paper, a data sample of proton-proton ($pp$) collisions at $\sqrt{s}=13\tev$, corresponding to
an integrated luminosity of $5\invfb$, collected by the LHCb experiment in years of 2016 to 2018, is used to 
study the decay of $\BtoLcLcK$. 
The $\Lc$ baryon is reconstructed in the $p \Km \pip$ final state.
The $\BtoDDK$ decay, with $\Dp\to \Km\pip\pip$, is chosen as the normalisation channel, since it has a very similar decay topology and the same number of tracks in the final state as the signal.
This allows a precise measurement of 
the ratio of the branching fractions of the $\BtoLcLcK$ and $\BtoDDK$ decays.

\section{Detector and signal selection}
\label{sec:selection}
The \lhcb detector~\cite{LHCb-DP-2008-001,LHCb-DP-2014-002} is a single-arm forward
spectrometer covering the \mbox{pseudorapidity} range $2<\eta <5$,
designed for the study of particles containing \bquark or \cquark
quarks. The detector includes a high-precision tracking system
consisting of a silicon-strip vertex detector (VELO) surrounding the $pp$
interaction region, a large-area silicon-strip detector located
upstream of a dipole magnet with a bending power of about
$4{\mathrm{\,Tm}}$, and three stations of silicon-strip detectors and straw
drift tubes placed downstream of the magnet.
The tracking system provides a measurement of the momentum, \ptot, of charged particles with
a relative uncertainty that varies from 0.5\% at low momentum to 1.0\% at 200\gev.
The minimum distance of a track to a primary $pp$ collision vertex (PV), the impact parameter (IP),
is measured with a resolution of $(15+29/\pt)\mum$,
where \pt is the component of the momentum transverse to the beam, in\,\gev.
The momentum scale is calibrated using samples of $\decay{\jpsi}{\mumu}$ and $\decay{\Bm}{\jpsi\Kp}$~decays collected concurrently with the data sample used for this analysis~\cite{LHCb-PAPER-2012-048,LHCb-PAPER-2013-011}. 
Different types of charged hadrons are distinguished using information
from two ring-imaging Cherenkov detectors.
Photons, electrons and hadrons are identified by a calorimeter system consisting of
scintillating-pad and preshower detectors, an electromagnetic
and a hadronic calorimeter. Muons are identified by a
system composed of alternating layers of iron and multiwire
proportional chambers.
The online event selection is performed by a trigger,
which consists of a hardware stage, based on information from the calorimeter and muon
systems, followed by a software stage, which applies a full event
reconstruction.
At the hardware trigger stage, events are required to have a muon with high \pt or a
hadron, photon or electron with high transverse energy in the calorimeters. 
The software trigger requires a two-, three- or four-track secondary vertex with a significant displacement from any primary $pp$ interaction vertex. 
Simulation is required to model the effects of the detector acceptance and the
imposed selection requirements.
In the simulation, $pp$ collisions are generated using
\pythia 8~\cite{Sjostrand:2007gs} 
with a specific \lhcb configuration~\cite{LHCb-PROC-2010-056}.
Decays of unstable particles are described by \evtgen~\cite{Lange:2001uf},
in which final-state radiation is generated using \photos~\cite{davidson2015photos}.
The interaction of the generated particles with the detector, and its response,
are implemented using the \geant toolkit~\cite{Allison:2006ve, *Agostinelli:2002hh}
as described in Ref.~\cite{LHCb-PROC-2011-006}. 

The $\Bm$ candidates are reconstructed with a pair of $\Lc\Lcbar$ baryons along with a companion kaon,
where the $\Lc$ baryon is reconstructed in the $\Lc \to p \Km \pip$ final state.
The $\Lc$ decay products are required to have good track quality and
large transverse momentum, to be inconsistent with originating from any PV,
and to have correct particle identification (PID) (proton, kaon or pion). 
The $\Lc$ candidate is required to have \pt greater than $1.8\gev$,
its vertex inconsistent with originating from any PV, and the angle
between its momentum and its flight direction with respect to the associated PV to be less than 90 degrees.
In case of multiple primary vertices, the candidate is associated to the PV with respect to which the smallest $\chisqip$ is obtained,
where $\chisqip$ is defined as the difference in the vertex-fit \chisq of the associated PV reconstructed with and without the track under consideration.
The reconstructed mass of the $\Lc$ candidates is required to be within the window of
$2225 < M(\proton\Km\pip) < 2345\mev$.
Higher thresholds on $\pt$ and momentum are set for the companion kaon
and similar requirements as for the $\Lc$ decay products are imposed
on its track quality, PID and displacement from the PV.
All the selection criteria for the $\Lc$ candidates apply to the $\Lcbar$ candidates.
The $\Bm$ candidates are required to have $\pt$ greater than $5\gev$,
a reconstructed lifetime longer than 0.2~ps, and to have a $\chisqip$ less than 25.
The reconstructed $\Bm$ flight direction, defined by its production and decay vertices,
and its momentum are required to be well aligned. The $\Bm$ candidate must be within the mass window of
$5205 < M(\Lc\Lcbar\Km) < 5355\mev$.

To further suppress the background, a boosted decision tree (BDT)~\cite{Breiman,AdaBoost},
implemented in the TMVA toolkit~\cite{Hocker:2007ht,*TMVA4}, is trained
with the simulated sample as signal and 
with data from sideband regions defined in Table~\ref{tab:sideband} as background.
The training variables are the $\chisq$ per degree of freedom of the $\Lc$ decay vertex,
and the $\chisqip$ and PID information of all final-state tracks.
The PID values in simulation are drawn from calibration samples
to better agree with real data~\cite{Aaij:2018vrk}. 
The selection requirement on the BDT output is chosen to optimise the expected significance
of the $\BtoLcLcK$ signal.

\begin{table}[t]
\begin{center}
\caption{\small Definitions of the sideband regions, where $m(\Bm)$ and $m(\Dp)$ represent the known values from Ref~\cite{PDG2020}. The same sideband region is defined for the $\antiproton\Kp\pim$ combination as for the $\proton\Km\pip$ combination. The values are in \mev.}
\begin{tabular}{lc} \hline
\multirow{3}{*}{$\BtoLcLcK$}    & $50< |M(\Lc\Lcbar\Km)-m(\Bm)| <90\mev$,\\
                                & $2225< M(\proton\Km\pip) < 2260\mev$ or \\
                                & $2310< M(\proton\Km\pip) < 2345\mev$ \\ \hline
\multirow{2}{*}{$\BtoDDK$}      & $75< |M(\Dm\Dp\Km)-m(\Bm)| <120\mev$,\\
                                & $45< |M(\Km\pip\pip)-m(\Dp)| <90\mev$      \\
\hline
\end{tabular}
\label{tab:sideband}
\end{center}
\end{table}

A dedicated BDT is trained for the normalisation channel $\BtoDDK$ 
with simulation as signal and
with candidates in the $\Dp$ and $\Bm$ sideband regions, as defined in Table~\ref{tab:sideband},
as the proxy for background.
The signal region of the $\BtoDDK$ decay is wider than the $\BtoLcLcK$ since
the energy release ($Q$-value) of the former is larger.
The selection is almost identical to
the signal selection, with the $\Lc$ candidate replaced by the $\Dp$ 
reconstructed from the $\Km\pip\pip$ decay mode.
The $\Dp$ candidates are required to be within a mass window of
$1795 < M(\Km\pip\pip) < 1945\mev$. 

\section{Signal extraction}
\label{sec:fit}
\begin{figure}
\begin{center}
\begin{minipage}[b]{0.48\textwidth}
\includegraphics[width=0.98\textwidth]{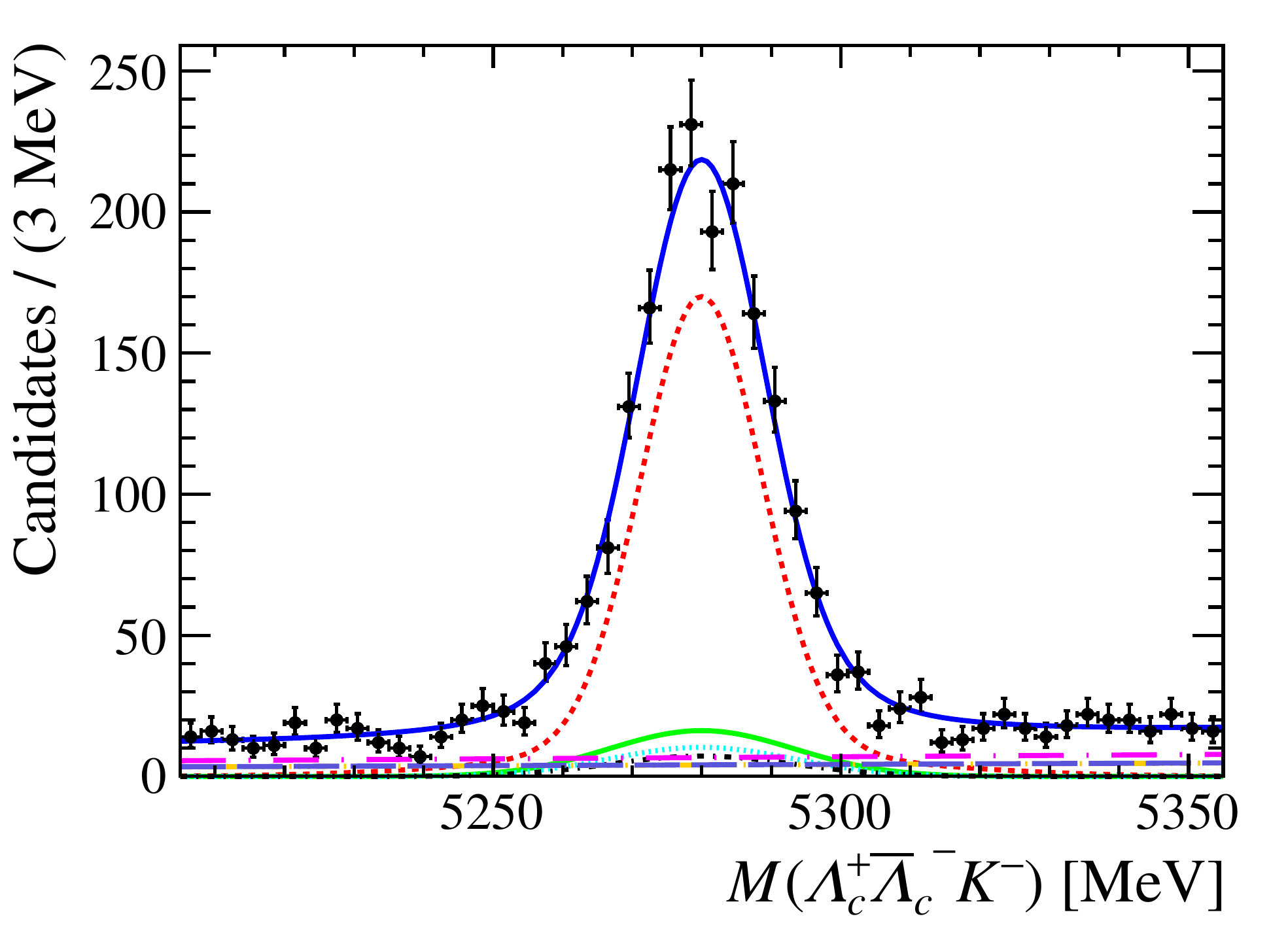}
\put(-170,125){LHCb}
\put(-170,110){5\,fb$^{-1}$}
\put(-35,125){(a)}
\end{minipage}
\begin{minipage}[b]{0.48\textwidth}
\includegraphics[width=0.98\textwidth]{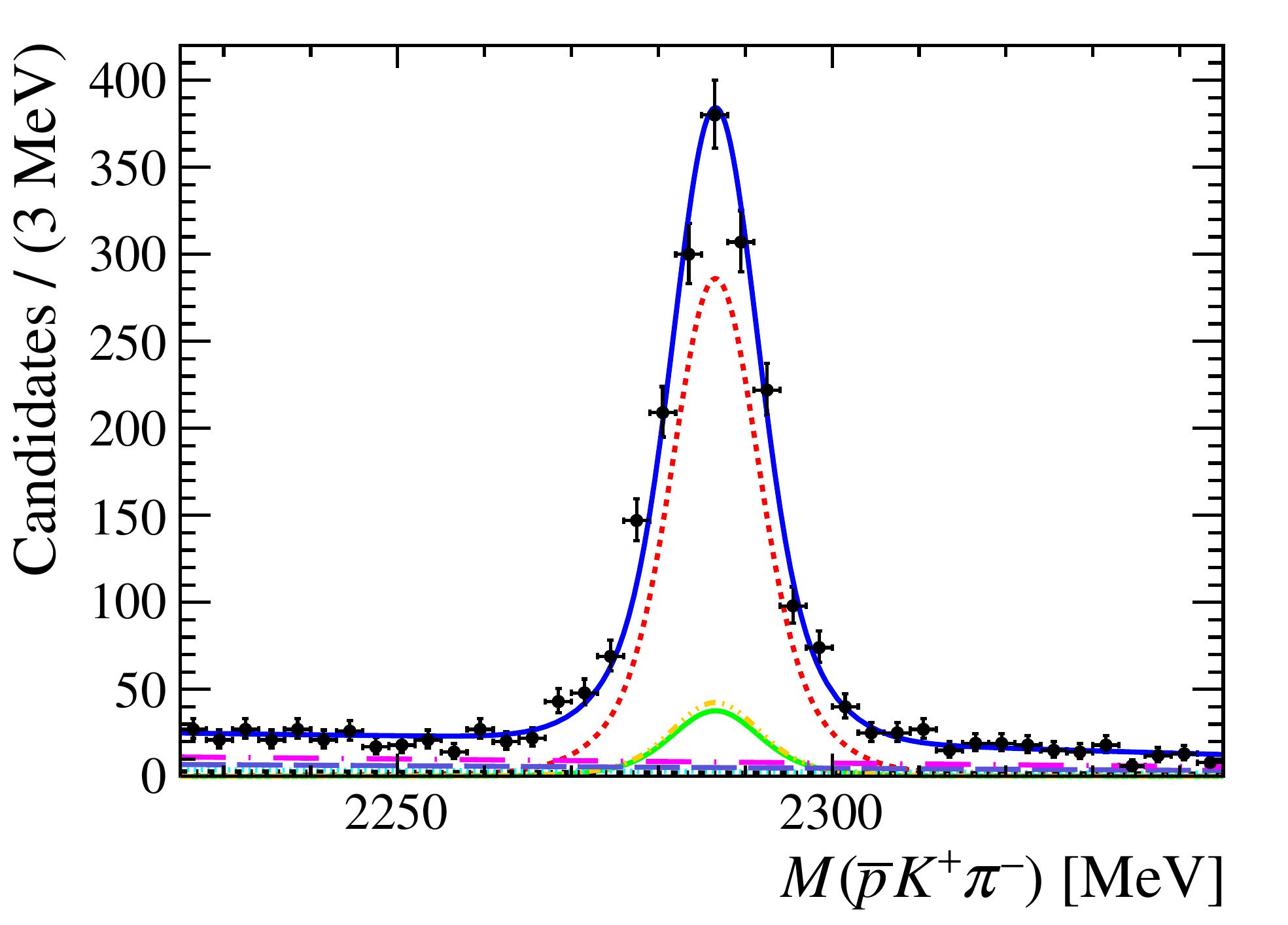}
\put(-170,125){LHCb}
\put(-170,110){5\,fb$^{-1}$}
\put(-35,125){(b)}
\end{minipage}
\begin{minipage}[b]{0.48\textwidth}
\includegraphics[width=0.98\textwidth]{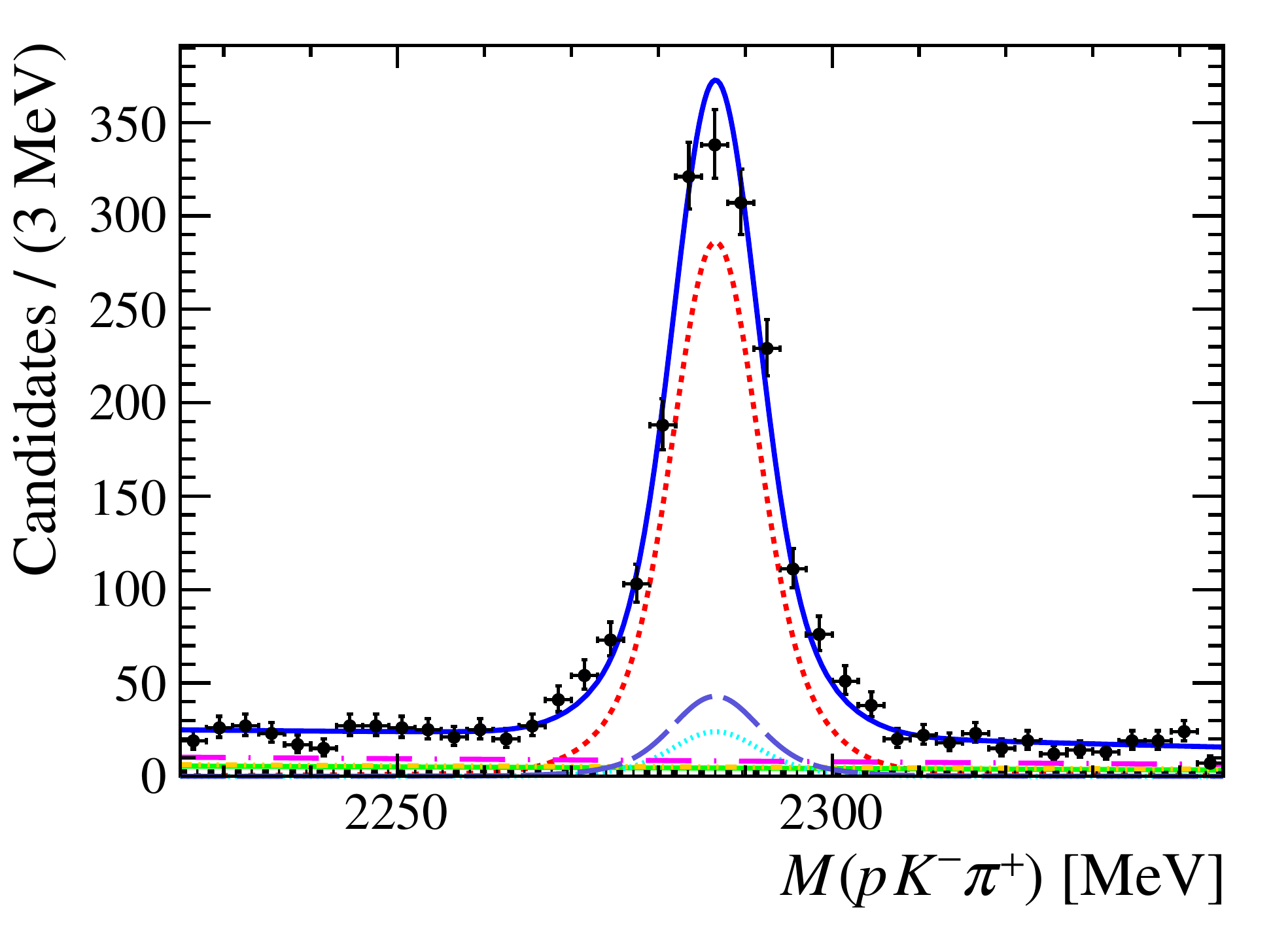}
\put(-170,125){LHCb}
\put(-170,110){5\,fb$^{-1}$}
\put(-35,125){(c)}
\end{minipage}
\begin{minipage}[b]{0.48\textwidth}
\includegraphics[width=0.80\textwidth]{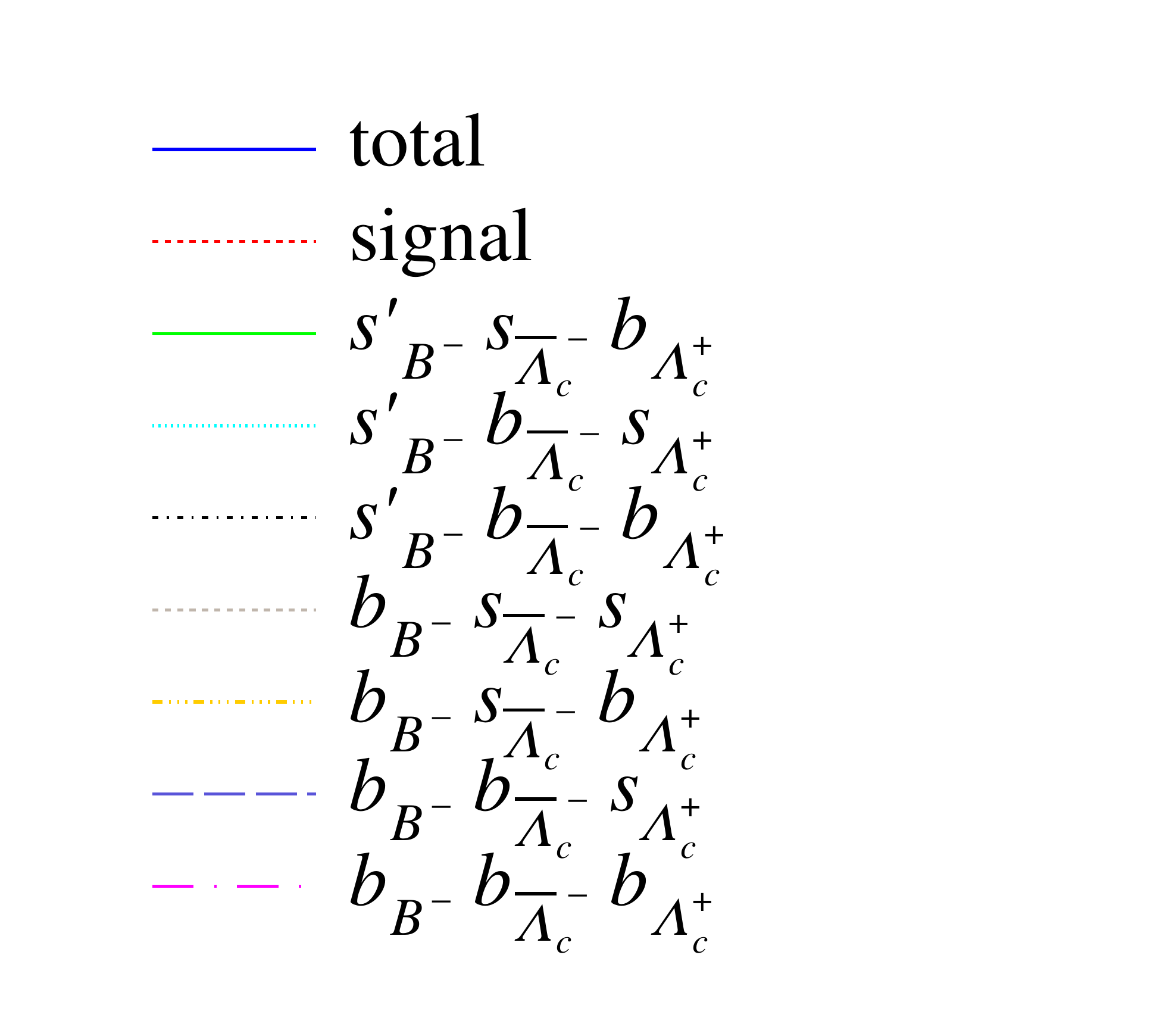}
\end{minipage}
\caption{\small Distributions of (a) $M(\Lc\Lcbar\Km)$, (b) $M(\antiproton\Kp\pim)$ and (c) $M(\proton\Km\pip)$ of selected $\BtoLcLcK$ candidates. The data points with error bars are shown along with the total fitted shape, which is composed of signal and background components, as shown in the legend.  }
\label{fig:Bu3Dfit}
\end{center}
\end{figure}

The $\BtoLcLcK$ signal is extracted using a three-dimensional (3D) fit to
the invariant mass of $\Bm$, $\Lc$ and $\Lcbar$ candidates.
The 3D fit function is constructed as a sum of multiple terms 
each of which is a direct product of one-dimensional functions.
The signal function is $s_{\Bm}s_{\Lcbar}s_{\Lc}$, where $s_{\Bm}$ is
modelled by sum of two Gaussian functions sharing common mean value and
$s_{\Lc}$ is modelled using sum of three Gaussian functions
with a common mean value fixed to the known $\Lc$ mass~\cite{PDG2020}.
The fit model and parameters are identical for $s_{\Lc}$ and $s_{\Lcbar}$.
The Gaussian parameters are all fixed to values from fits to simulated samples.
The background shape in the $M(\Lc\Lcbar\Km)$ and $M(\proton\Km\pip)$ distributions is parameterised with a first-order polynomial.
The background candidates are of seven different categories:
from real $\Bm$ decay products, (1) $s^{\prime}_{\Bm}s_{\Lcbar}b_{\Lc}$,
(2) $s^{\prime}_{\Bm}b_{\Lcbar}s_{\Lc}$, (3) $s^{\prime}_{\Bm}b_{\Lcbar}b_{\Lc}$;
from non-$\Bm$ decays containing true $\Lz_{\cquark}$ decays, (4) $b_{\Bm}s_{\Lcbar}s_{\Lc}$,
(5) $b_{\Bm}s_{\Lcbar}b_{\Lc}$, (6) $b_{\Bm}b_{\Lcbar}s_{\Lc}$;
and pure combinatorial background, (7) $b_{\Bm}b_{\Lcbar}b_{\Lc}$.
For background categories (1)--(3) the $Q$-values of $\Bm$ decays could
differ from the signal, and a single Gaussian $s^{\prime}_{\Bm}$ 
is used.
The fitted $\Bm$, $\Lc$ and $\Lcbar$ mass spectra are shown
in Fig.~\ref{fig:Bu3Dfit}. The signal yield is determined to be $1365 \pm 42$.
It is found that background (4) $b_{\Bm}s_{\Lcbar}s_{\Lc}$ makes negligible contribution; thus the yield is fixed to zero.

\begin{figure}[htbp]
\begin{center}
\begin{minipage}[b]{0.50\textwidth}
\includegraphics[width=0.98\textwidth]{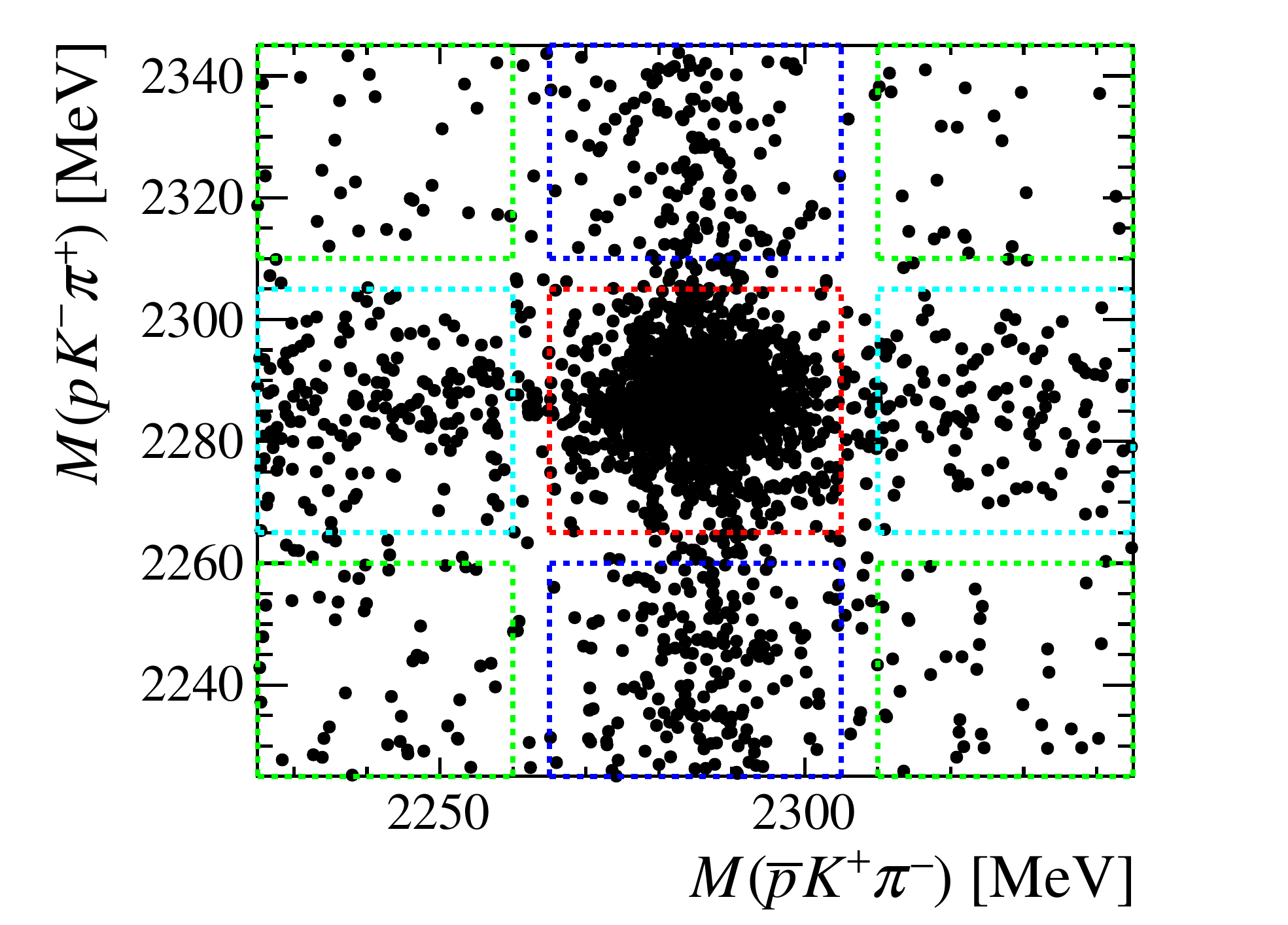}
\end{minipage}
\caption{\small Selected $\BtoLcLcK$ candidates in the two-dimensional $M(\proton\Km\pip)$ and $M(\antiproton\Kp\pim)$ spectrum. The red dashed box is the signal region, the dark blue, light blue and green boxes indicate regions dominated by $s_{\Lcbar}b_{\Lc}$, $b_{\Lcbar}s_{\Lc}$ and combinatorial background, respectively. }
\label{fig:regions2D}
\end{center}
\end{figure}

The resonance structure present in the $\BtoLcLcK$ decay is studied using the candidates
in narrower $\Bm$ and $\Lc$ mass windows:
$5240 < M(\Lc\Lcbar\Km) < 5320\mev$ and \mbox{$2265 < M(\proton\Km\pip) < 2305\mev$}, respectively.
In order to improve the mass resolution of any resonant state,
a refit of the decay fixing the masses of
$\Lc$ and $\Bm$ to the known values is performed.
The refitted momenta of $\Lc$, $\Lcbar$ and $\Km$ are used to determine the $M(\Lc\Km)$,
$M(\Lcbar \Km)$ and $M(\Lc \Lcbar)$ mass spectra.
It is verified that the selection requirements do not induce artificial peaking structures.
For background subtraction, it is sufficient to consider the sideband in the two-dimensional
spectrum of $M(\proton\Km\pip)$ and $M(\antiproton\Kp\pim)$, as shown in Fig.~\ref{fig:regions2D},
since the absence of the $b_{\Bm}s_{\Lcbar}s_{\Lc}$ background component means that
whenever a true $\Lc\Lcbar$ pair is selected, it corresponds
to a real $\BtoLcLcK$ decay.
The contributions from $s_{\Lcbar}b_{\Lc}$ and $b_{\Lcbar}s_{\Lc}$ are averaged to
account for the background contribution in the signal region, where the double-counted
contribution from pure combinatorial background, as in $b_{\Lcbar}b_{\Lc}$, is subtracted.
The background subtraction is also performed with the $sPlot$ 
method~\cite{Pivk:2004ty} as a cross-check
to verify that none of the observed structures are due to background fluctuations.

\begin{figure}[htbp]
\begin{center}
\begin{minipage}[b]{0.60\textwidth}
\includegraphics[width=0.98\textwidth]{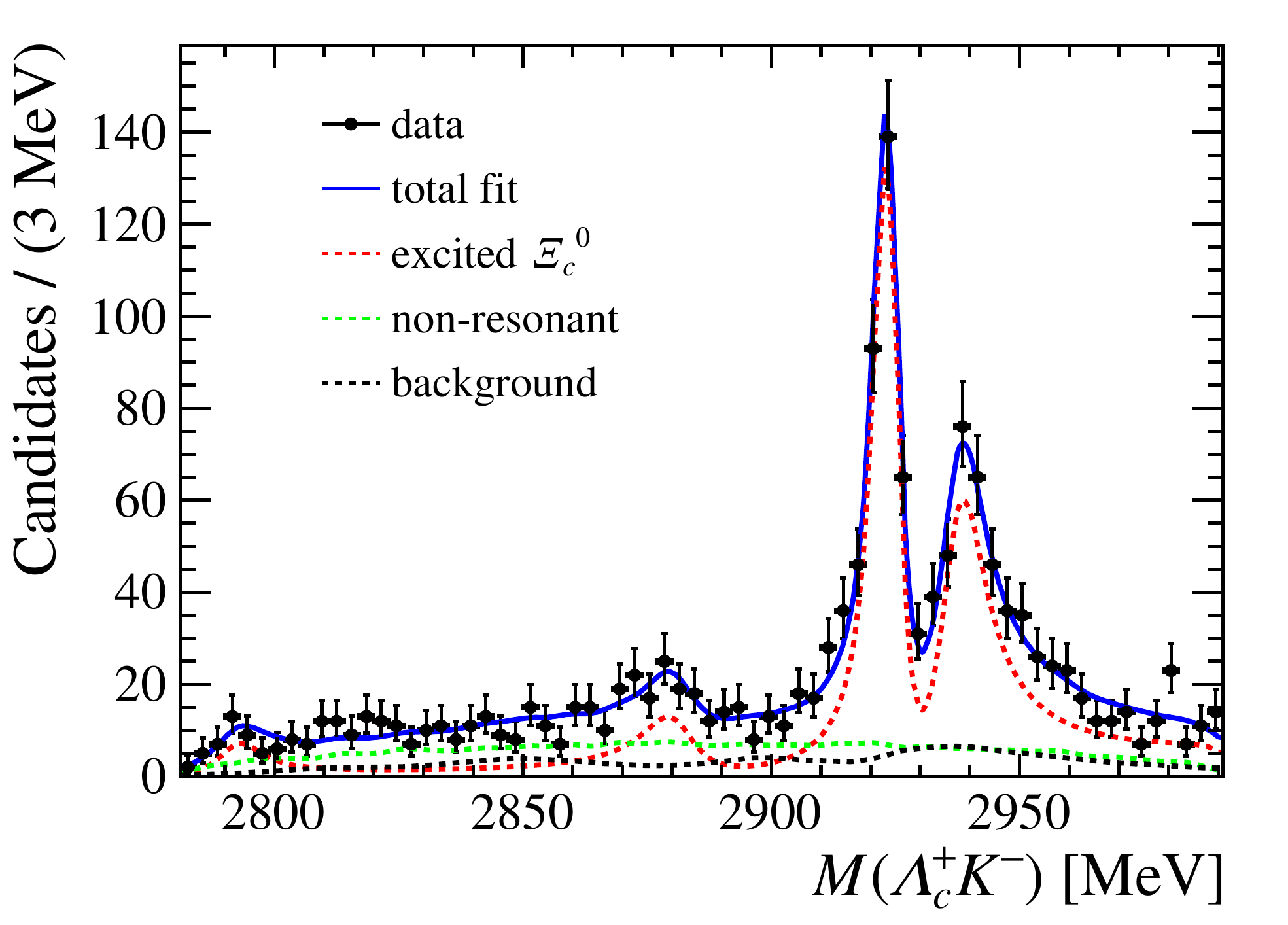} 
\put(-60,160){LHCb}
\put(-60,145){5\,fb$^{-1}$}
\end{minipage}
\caption{\small Mass spectrum of the $\Lc\Km$ pair from the $\BtoLcLcK$ decays. The data points with error bars are shown along with the total fitted shape in blue solid line, which is composed of the components, as shown in the legend. }
\label{fig:MLcbarKp}
\end{center}
\end{figure}

The $M(\Lc\Km)$ spectrum is shown in Fig.~\ref{fig:MLcbarKp}.
The fitting function is described in detail in Sec.~\ref{sec:Xic-fit}.
Two states reported in Ref.~\cite{LHCb-PAPER-2020-004}, $\Xic(2923)^{0}$ and $\Xic(2939)^{0}$, are observed.
In addition, a wide structure is evident around $2880\mev$.
A similar structure was observed in the prompt $\Lc\Km$ study~\cite{LHCb-PAPER-2020-004},
but background due to feed-down from higher $\Xicz$ states, $\Xic(3055)^{0}$ and
$\Xic(3080)^{0}$, with a $\piz$ or $\pip$ missing in the reconstruction,
could not be discounted.
Such background will not affect the $\Lc\Km$ system from $\Bm$ decays, since there is not enough phase space for $\Bm \to \Lcbar\Xic(3055/3080)^{0}$ and
such feed-down component with a missing particle will not peak at the $\Bm$ mass.
To study the effect of partially reconstructed $\Xic(3055/3080)^{0}$ from other $b$-hadron decays,
a sample of $\Lb$ decays is generated 
with the decay chain
\mbox{$\Lb \to \Xic(3055)^{+} \Dm$, $\Xic(3055)^{+} \to [\Lc \pip]_{\Sigmares_\cquark^{++}}\Km,
\Dm \to \Kp\pim\pim$}.
The selection criteria remove them completely.
Therefore, the wide structure cannot be due to feed-down, and is considered
as a resonant state in the fit.
Both $\Xic(2790)^{0}$ and $\Xic(2815)^{0}$ states are expected at the lower end of the $M(\Lc\Km)$ spectrum,
and their masses, widths and spin-parity $J^{P}$ are fixed to known values~\cite{PDG2020}. 
The $\Xic(2790)^{0}$ state has $J^{P} = (1/2)^{-}$ and $\Xic(2815)^{0}$ has $J^{P} = (3/2)^{-}$.
However, the significance of the $\Xic(2815)^{0}$ state is \mbox{only $2.1\,\sigma$}, and so it is
not included in the nominal fit.
The yield of the resonant states is not enough to determine their quantum numbers with sufficient significance. 
The spin-parity $J^{P}$ of both $\Xic(2923)^{0}$ and $\Xic(2939)^{0}$ states is fixed to be
$(3/2)^-$ and that of $\Xic(2880)^{0}$ is fixed to $(1/2)^-$~\cite{Wang:2020gkn}.

\begin{figure}[htbp]
\begin{center}
\begin{minipage}[b]{0.48\textwidth}
\includegraphics[width=0.98\textwidth]{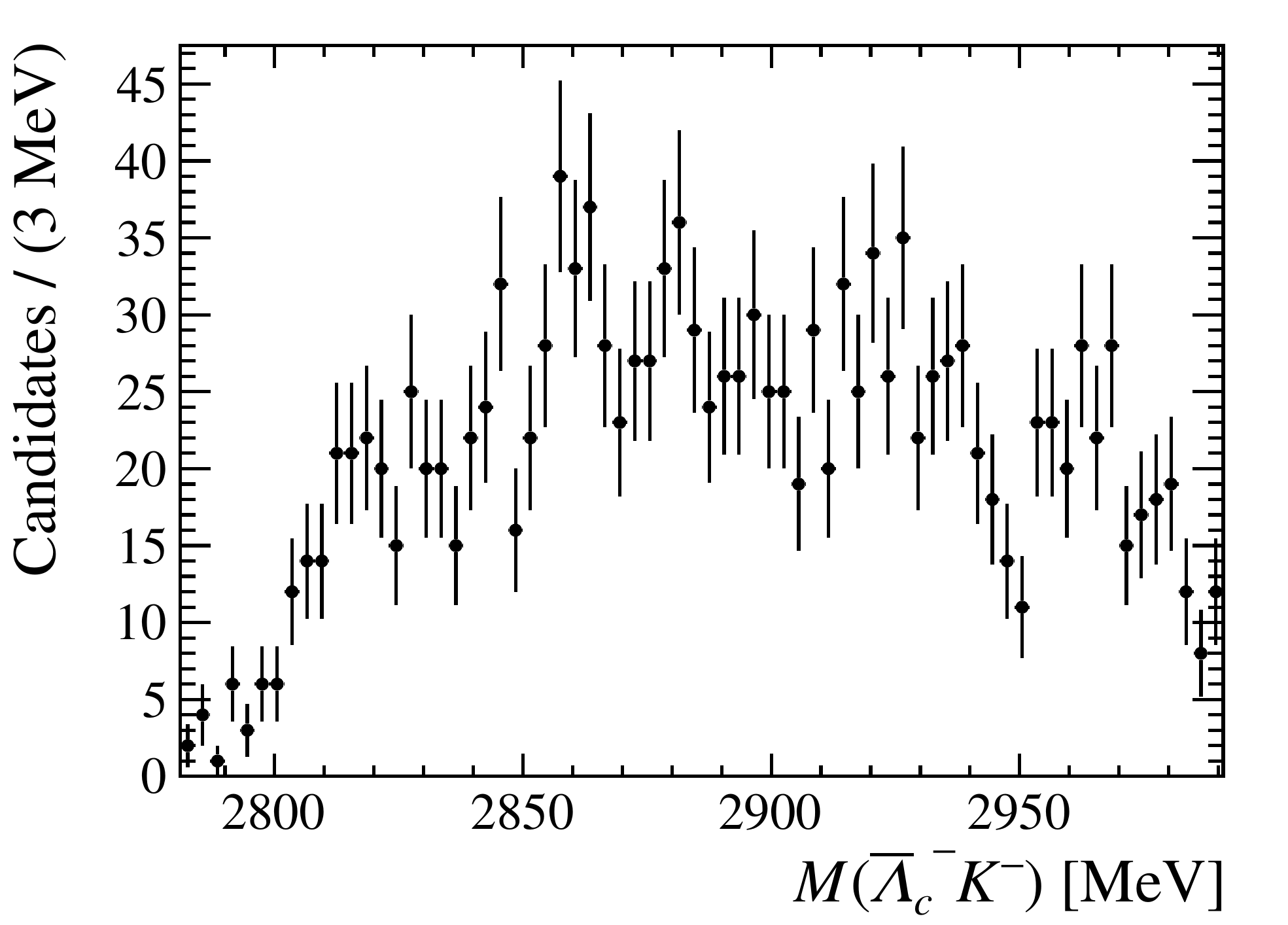}
\put(-175,130){LHCb}
\put(-175,115){5\,fb$^{-1}$}
\put(-35,130){(a)}
\end{minipage}
\begin{minipage}[b]{0.48\textwidth}
\includegraphics[width=0.98\textwidth]{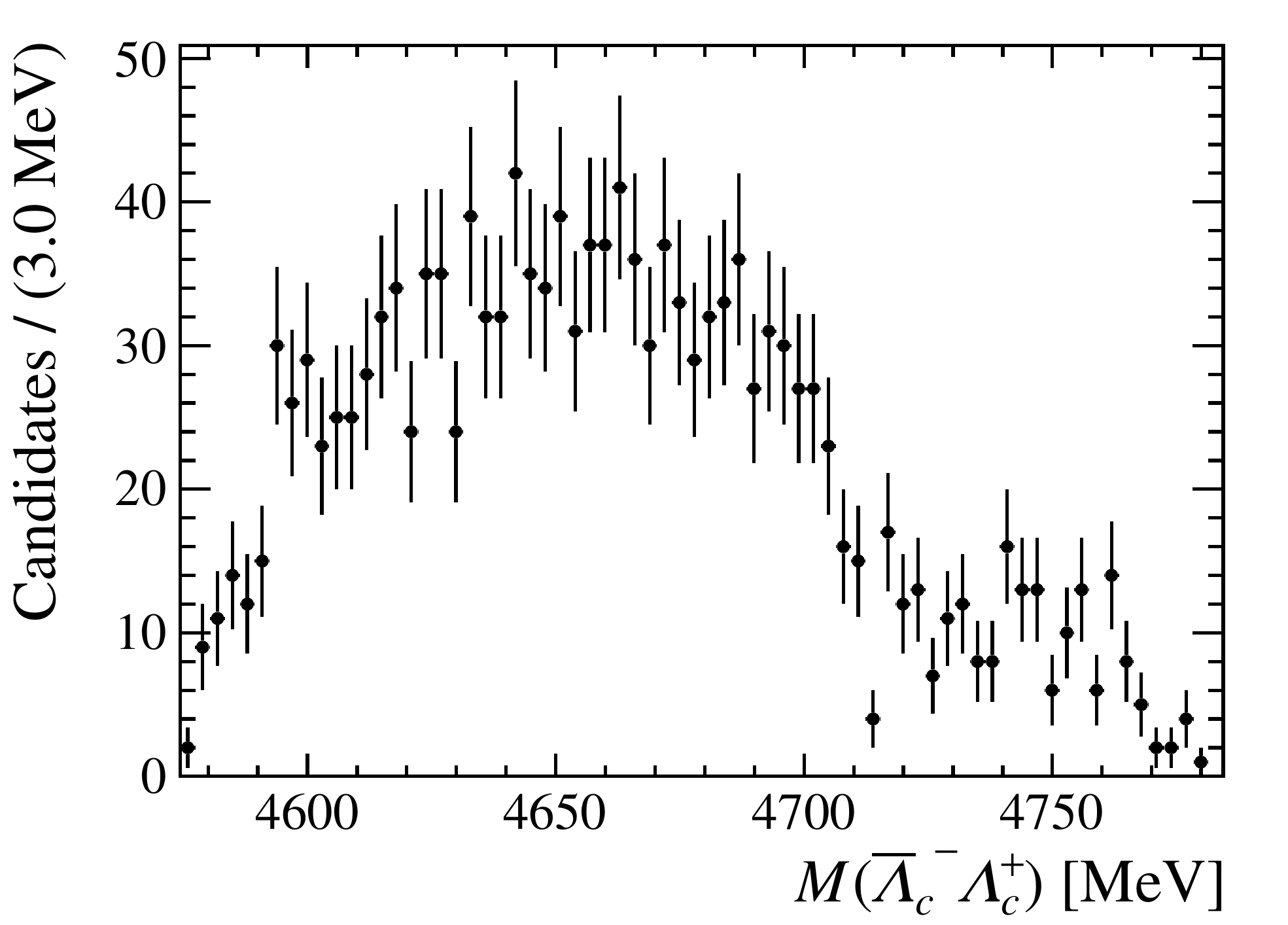}
\put(-55,130){LHCb}
\put(-55,115){5\,fb$^{-1}$}
\put(-175,130){(b)}
\end{minipage}
\begin{minipage}[b]{0.48\textwidth}
\includegraphics[width=0.98\textwidth]{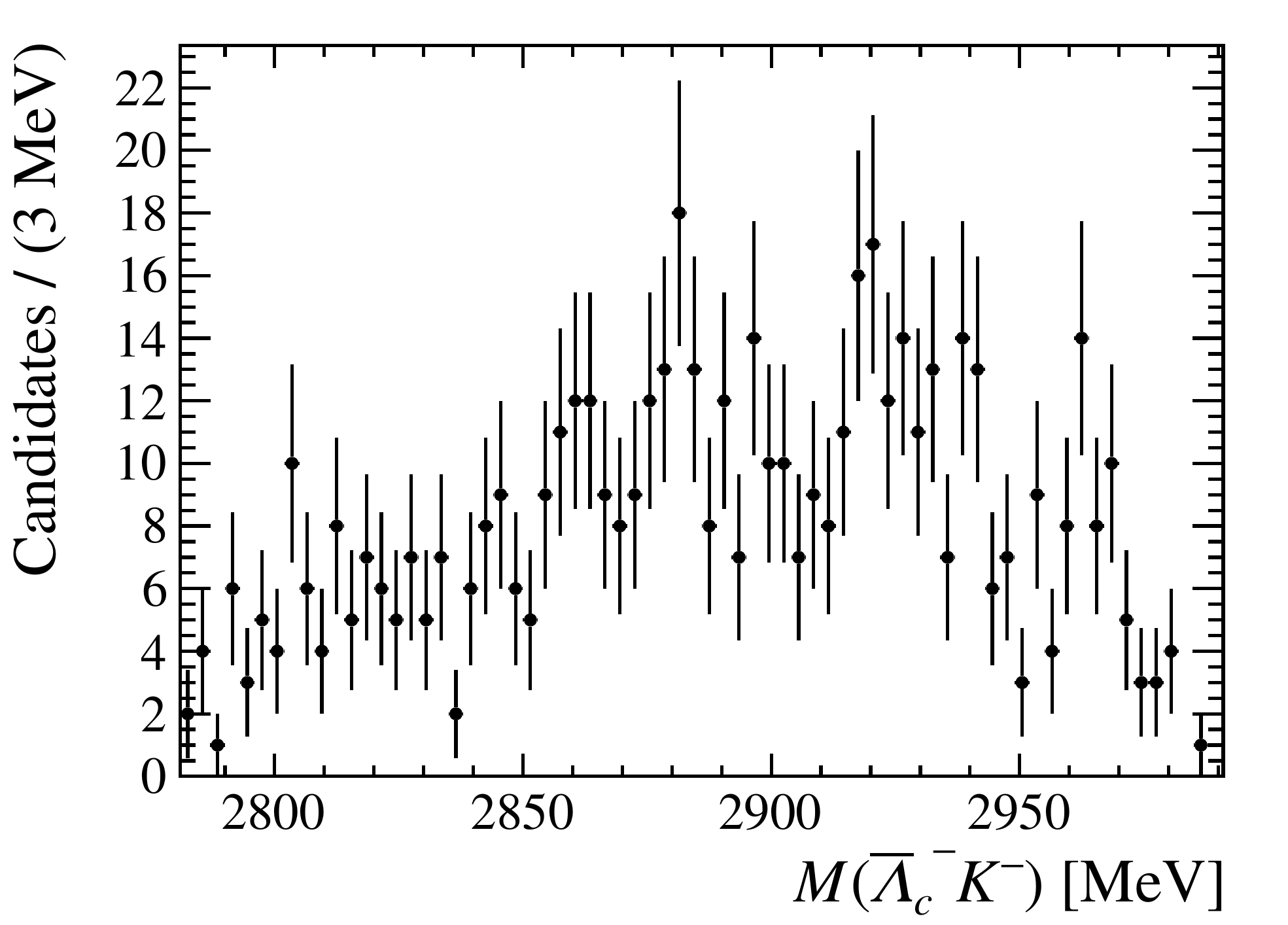}
\put(-175,130){LHCb}
\put(-175,115){5\,fb$^{-1}$}
\put(-35,130){(c)}
\end{minipage}
\begin{minipage}[b]{0.48\textwidth}
\includegraphics[width=0.98\textwidth]{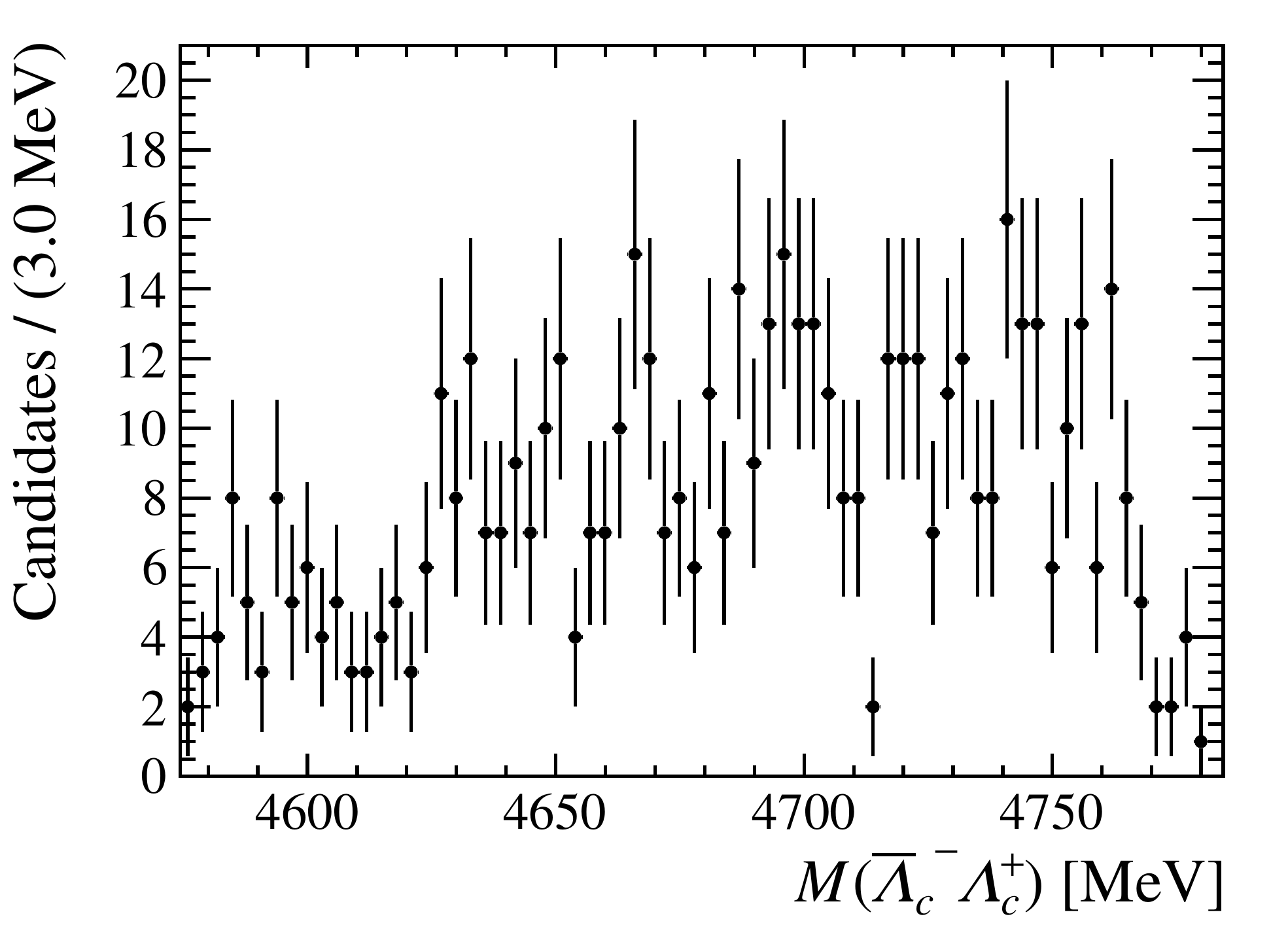}
\put(-175,130){LHCb}
\put(-175,115){5\,fb$^{-1}$}
\put(-70,130){(d)}
\end{minipage}
\caption{\small Mass spectrum of (a,c) $\Lcbar\Km$ and (b,d) $\Lc\Lcbar$ pairs from (a,b) all the $\BtoLcLcK$ candidates and (c,d) after vetoing candidates with $2900 < M(\Lc\Km) < 2970\mev$.}
\label{fig:MLcLc}
\end{center}
\end{figure}

The invariant mass distributions of the $\Lcbar\Km$ and $\Lc\Lcbar$ pairs are shown in Fig.~\ref{fig:MLcLc}. 
No significant structure is seen.
The distributions of $M(\Lcbar \Km)$ and $M(\Lc\Lcbar)$ are also produced after rejecting candidates
with \mbox{$2900 < M(\Lc\Km) < 2970\mev$}, to remove contributions from the $\Xic(2923)^{0}$ and
$\Xic(2939)^{0}$ states. No significant structure is seen.

\section{Masses and widths of the excited $\Xicz$ states}
\label{sec:Xic-fit}
The masses and widths of the excited $\Xicz$ states are obtained from an unbinned maximum
likelihood fit to the $M(\Lc\Km)$ spectrum.
The total fitting function is constructed as
\begin{equation}
   f(M) = \frac{N_{\Xicz}}{\nevt}\,f_{\Xicz}(M)\otimes g(\sigma_{\mathrm{res}})
  + \frac{N_{\phsp}}{\nevt}\, f_{\phsp}(M)
  + \frac{N_{\bkg}}{\nevt}\,f_{\bkg}(M),   
  \label{eq:totalpdf}
\end{equation}
where $M$ stands for $M(\Lc\Km)$. 
The function describing the excited $\Xicz$ states, $f_{\Xicz}$,
which will be discussed in detail later, is convolved with a Gaussian function
to account for the detector resolution. The Gaussian function has a mean value of zero
and width $\sigma_{\mathrm{res}}$ in the range of 0--2\mev dependent on $M(\Lc\Km)$. The resolution $\sigma_{\mathrm{res}}$ is parameterised as \mbox{$(M-m_{\mathrm{lo}})^a(m_{\mathrm{up}}-M)^b[c+d(M-m_{\mathrm{lo}})]$},
where \mbox{$m_{\mathrm{lo}}=m(\Lc)+m(\Km)$} and \mbox{$m_{\mathrm{up}}=m(\Bm)-m(\Lcbar)$} are lower and upper thresholds of the $M(\Lc\Km)$ spectrum, respectively, and $a, b, c, d$ are determined from simulation.
The contribution from the non-resonant component $f_{\phsp}$ is modelled with simulation of pure phase-space decay.
The background contribution $f_{\bkg}$ is extrapolated from the sideband regions as defined in
Fig.~\ref{fig:regions2D}. The expected number of background events $N_{\bkg}$
is also fixed using extrapolation from the sideband regions.
The total number of candidates within the fitting range, $\nevt$, is 1494.
All fitting functions are normalised to unity.

As discussed in Sec.~\ref{sec:fit}, four excited states are considered:
$\Xic(2790)^{0}$, $\Xic(2880)^{0}$, $\Xic(2923)^{0}$ and $\Xic(2939)^{0}$.
The interference between the $\Xic(2923)^{0}$ and $\Xic(2939)^{0}$ states is 
important and cannot be neglected to correctly describe the data.
The significance of the two-state hypothesis with respect to the hypothesis of a single $\Xic(2930)^{0}$ state is over $11\,\sigma$.
The function used for excited $\Xicz$ states is expressed as
\begin{align}
f_{\Xicz} = p\,q \left(\left|\mathcal{M}_{\Xic(2790)^0} + \mathcal{M}_{\Xic(2880)^0}\right|^2 + \left|\mathcal{M}_{\Xic(2923)^0} + \mathcal{M}_{\Xic(2939)^0}\right|^2\right) \, ,
\end{align}
where $\mathcal{M}_{\alpha}(M) = \hat{A}_{\alpha}(M)F_{l_{\alpha}}(q)F_{L_{\alpha}}(p)$.
The quantities $q$ and $p$ represent the breakup momenta of the $\Xicz$ and $\Bm$ decays, expressed as 
\mbox{$q = \lambda^{1/2}(M^2,m_K^2,m_{\Lambda_c}^2)/(2M)$},
\mbox{$p = \lambda^{1/2}(m_B^2,M^2,m_{\Lambda_c}^2)/(2M)$}, where
\mbox{$\lambda(x,y,z) = x^2+y^2+z^2-2xy-2yz-2zx$}.
The \mbox{subscript $\alpha$} runs through all four resonant states considered in the fit.
The amplitude of each state is described by a relativistic Breit--Wigner function,
\begin{align}
  \hat{A}_{\alpha}(M)
  = \frac{c_{\alpha}}{m_{\alpha}^2-M^2-i\,m_{\alpha} \,\Gamma_{\alpha}\,\frac{q}{q_{\alpha}}\cdot\frac{m_{\alpha}}{M}\cdot\frac{F_{l_{\alpha}}(q)^2}{F_{l_{\alpha}}(q_{\alpha})^2}},
\end{align}
where $m_{\alpha}$ and $\Gamma_{\alpha}$ are the mass and width of the state,
$c_{\alpha}$ is a complex coefficient, and $q_{\alpha}$ is the breakup momentum $q$ computed at the mass $m_{\alpha}$.
The quantities $F_{l}(q)$ and $F_{L}(p)$ are Blatt--Weisskopf barrier functions, defined as
\begin{align}
F_{k}(q) = \begin{cases} 1                           &k = 0, \\ 
\sqrt{\frac{(qr)^{2}}{1+(qr)^{2}}}                   &k = 1, \\ 
\sqrt{\frac{(qr)^{4} }{9 + 3(qr)^{2} + (qr)^{4} } }  &k = 2,
\end{cases}          
\end{align}
where $r$ is the effective radius of the resonant state, fixed to $3.0\gev^{-1}$~\cite{LHCb-PAPER-2016-019},
$l$ is the orbital angular momentum between the $\Lcbar$ and $\Kp$ particles, and
$L$ is the smallest possible orbital angular momentum between the $\Lc$ and $\Xicz$ states.
As mentioned before, the spin-parity of the $\Xic(2790)^{0}$ state is known to be $(1/2)^{-}$~\cite{PDG2020}.
The 1P multiplet contains 5 states with the quantum numbers $(1/2)^-, (1/2)^-, (3/2)^-, (3/2)^-$ and $(5/2)^-$.
The quantum numbers of the $\Xic(2880)^{0}$ state are assumed to be $(1/2)^{-}$ as well~\cite{Wang:2020gkn}.
The spin-parity of the $\Xic(2923)^{0}$ and $\Xic(2939)^{0}$ states is assumed to be $(3/2)^{-}$.
The values of $l$ are set to respect momentum and parity conservation,
namely $l=0$ for the $\Xic(2790)^{0}$ and $\Xic(2880)^{0}$ states 
and $l=2$ for the $\Xic(2923)^{0}$ and $\Xic(2939)^{0}$ states.
Alternative assumptions for the quantum numbers of the states are considered in the determination of the systematic uncertainty.
If the spin of the $\Xicz$ of interest is $J$, then $L$ can only take values
$J-1/2$ and $J+1/2$, and $F_{L_{\alpha}}$ is replaced by
$F_{J_{\alpha}-1/2} + k_{\alpha}F_{J_{\alpha}+1/2}$.
The $k_{\alpha}$ parameters are complex factors representing the contribution of the higher partial waves.
They cannot be extracted from a one-dimensional fit with the available
statistics, and all $k_{\alpha}$ parameters are set to a common value of $k_{0}$ in the fit. 

To study the possible bias on the measured mass and width of the $\Xic(2923)^{0}$ and $\Xic(2939)^{0}$ states,
$3\,000$ pseudoexperiments are performed where all other parameters, except the masses and widths, are fixed.
A fit is performed for each pseudoexperiment, and the pull of each mass or width parameter
is calculated with respect to the input.
The pull is defined as the difference between the fitted value and the input value, divided by the uncertainty obtained from the fit.
The pull distributions are then fitted with Gaussian functions. The deviation of the Gaussian mean 
from zero is used to correct the fitted mass values. 
The correction values are smaller than the statistical uncertainties and will be considered in the systematic uncertainty determination.

The fitted $M(\Lc\Km)$ distribution is shown in Fig.~\ref{fig:MLcbarKp}, and the measured masses and widths are listed in Table~\ref{tab:2Dfitresstat}.
The significance of the $\Xic(2790)^{0}$ and $\Xic(2880)^{0}$ states is calculated by
studying $30\,000$ pseudoexperiments. Each is generated with a null hypothesis, then fitted
both with and without the excited $\Xicz$ state of interest. The test statistic $t_0$,
defined as twice the difference in log-likelihood with and without the state,
$2\log(\mathcal{L}_1/\mathcal{L}_0)$, is expected to follow a $\chisq$ distribution.
The $t_0$ values from the pseudoexperiments are fitted with a $\chisq$ distribution, and
the $p$-value of the observed yield corresponds to the fraction of integrated area
above the $t_{0}$ value measured in real data divided by the total integrated area.
The significance of both the $\Xic(2790)^{0}$ and $\Xic(2880)^{0}$ states is estimated to be
$3.9\,\sigma$. The significance of the $\Xic(2880)^{0}$ state is stable even assuming the absence of the $\Xic(2790)^{0}$ state.

\begin{table}[t]
\begin{center}
\caption{\small Measured masses, widths and significance of excited $\Xicz$ states. }
\begin{tabular}{lccc} \hline
  State           & Mass (MeV)         & Width (MeV) & Significance  \\ \hline
$\Xic(2880)^{0}$      & $2881.8 \pm 3.1 \pm 8.5$   & $12.4 \pm 5.2 \pm 5.8$  & $3.8 \sigma$ \\  
$\Xic(2923)^{0}$      & $2924.5 \pm 0.4 \pm 1.1$   & $\phantom{0}4.8 \pm 0.9 \pm 1.5$ & $> 10 \sigma$ \\
$\Xic(2939)^{0}$      & $2938.5 \pm 0.9 \pm 2.3$   & $11.0 \pm 1.9 \pm 7.5$   & $> 10 \sigma$ \\
\hline
\end{tabular}
\label{tab:2Dfitresstat}
\end{center}
\end{table}

Systematic uncertainties on the mass and width measurements from various sources are studied.
Multiple alternative assumptions on the fixed parameters are tested.
The spin-parity of the $\Xic(2923)^{0}$ and $\Xic(2939)^{0}$ states is set to $(1/2)^-$,
$(1/2)^+$ or $(3/2)^-$, and $J^{P}$ of the $\Xic(2880)^{0}$ is set to $(1/2)^+$, $(3/2)^-$ or $(3/2)^+$. 
In these tests, the states with the same spin-parity are always added coherently.
The effective radius $r$ is set to either 2.0 or $4.0\gev^{-1}$.
The mass and width of the $\Xic(2790)^0$ state are varied within their uncertainty, 
and the hypothesis without the $\Xic(2790)^0$ state is tested.
A different coefficient $k_{\alpha}$ is assigned to each group of $\Xicz$ states with the same spin-parity.
An additional state around 2970\mev with orbital angular momentum of 0, 1 or 2 is added.
The fit including the $\Xic(2815)^{0}$ state is considered.
The potential interference with non-resonant decays is considered by adding a constant
term in the $\Lc\Km$ mass distribution.
The maximum variation in the fit results is obtained for each alternative assumption if multiple values are considered, and the total systematic uncertainty due to model assumptions is the quadratic sum of variation in all alternative assumptions.

The $\Xic(2923)^{0}$ and $\Xic(2939)^{0}$ lineshape is described alternatively using a K-matrix formalism~\cite{Aitchison:1972ay}, which preserves unitarity. 
The variation in the results is considered as a systematic uncertainty.
The pull distributions from pseudoexperiments are used to correct a possible bias
in the masses and widths of the $\Xic(2923)^{0}$ and $\Xic(2939)^{0}$ states, 
where the resulting corrections are smaller than statistical uncertainties. 
It is found that the fitted width $\sigma_{\rm{pull}}$ of the pull distribution is 
slightly wider than unity, indicating a potential under-estimation of the statistical uncertainty. 
Therefore, the term $\sqrt{{\sigma}^2_{\rm{pull}}-1}\cdot\sigma_{\mathrm{stat}}$ 
is assigned as a systematic uncertainty.
A relative uncertainty of $3 \times 10^{-4}$ on the charged particle momentum scale~\cite{LHCb-PAPER-2020-004} is propagated through the simulation. 
The resolution $\sigma_{\mathrm{res}}$ is varied within uncertainty and
the resulting difference on the fitted parameters of the resolution function is
then propagated to the masses and natural widths of the excited $\Xicz$ states,
which is found to be negligible.
The magnitude of the energy-loss correction for charged particles is known to $10$\% accuracy~\cite{LHCb-PAPER-2010-001}.
In the study of Ref.~\cite{LHCb-PAPER-2013-011}, a correction of less than $0.01\mev$ per track is estimated.
A conservative estimation of $0.01\mev$ per track is taken as the systematic uncertainty,
which is also found to be negligible.

Mass constraints on the $\Bm$ and $\Lc$ candidates are applied when calculating $M(\Lc\Km)$.
The maximum change after varying the $\Bm$ or $\Lc$ mass by one standard deviation
is assigned as a systematic uncertainty.
The systematic uncertainty due to the background is estimated by generating and fitting
300 pseudoexperiments, randomly varying the background yield according to a
Poisson distribution while fixing the $\Xicz$ yield, the parameters of the $\Xicz$ model
and the non-resonant yield. The average of the fitted values is calculated,
and its difference from the nominal result is assigned as a systematic uncertainty.

A summary of the systematic uncertainties are listed in
Table~\ref{tab:sysmasswidth}. The uncertainties from different sources are uncorrelated and are added in quadrature to give the total systematic uncertainty.
The significance of the $\Xic(2880)^{0}$ and $\Xic(2790)^{0}$ states is $3.8\,\sigma$ and $3.7\,\sigma$,
respectively, after taking the systematic uncertainties into account.

\begin{table}[t]
\begin{center}
\caption{\small Systematic uncertainties on the masses and widths of the $\Xic(2923)^{0}$, $\Xic(2939)^{0}$ and $\Xic(2880)^{0}$ states. Values are given in $\mev$.}
\begin{tabular}{lcccccc} \hline
 & \multicolumn{2}{c}{$\Xic(2923)^{0}$}      & \multicolumn{2}{c}{$\Xic(2939)^{0}$}               & \multicolumn{2}{c}{$\Xic(2880)^{0}$}      \\ \hline
Source                          & Mass  & Width & Mass  & Width & Mass  & Width \\ \hline
Model assumption                & 0.8   & 1.4   & 1.9   & 7.0   & 8.4   & 4.1   \\
Lineshape formalism            & 0.7   & 0.1   & 1.3   & 0.8   & 1.1   & 0.0   \\
Bias correction  & 0.2   & 0.4   & 0.4   & 2.1   & $-$   & $-$   \\       
Momentum scale                & 0.0   & $-$   & 0.1   & $-$   & 0.0   & $-$   \\
Mass constraint                 & 0.1   & 0.1   & 0.1   & 0.5   & 0.1   & 0.4   \\ 
Background                      & 0.1   & 0.3   & 0.2   & 0.8   & 0.7   & 4.0   \\ \hline
Total                           & 1.1   & 1.5   & 2.3   & 7.5   & 8.5   & 5.8   \\
\hline
\end{tabular}
\label{tab:sysmasswidth}
\end{center}
\end{table}

\section{Branching fraction}
\label{sec:BF}
The relative branching fraction $R_{\BF}$ of the $\BtoLcLcK$ decay with respect to the $\BtoDDK$ channel
is calculated as a ratio of efficiency-corrected yield divided by
the corresponding charm decay branching fractions, expressed as
\begin{align}
    R_{\BF} = \frac{N(\BtoLcLcK)}{N(\BtoDDK)}
     \times \frac{\etot(\BtoDDK)}{\etot(\BtoLcLcK)}
     \times \frac{\BF^{2}(\Dp \to \Km \pip \pip)}{\BF^{2}(\Lc \to p \Km \pip)}.
\end{align}
The yield of the signal and normalisation channels is obtained from 
the 3D fit described in Sec.~\ref{sec:fit}.
The charm decay branching fractions $\BF(\Dp \to \Km \pip \pip)$ [$(9.38 \pm 0.16)$\%] and $\BF(\Lc \to p \Km \pip)$ [$(6.28 \pm 0.32)$\%] are obtained from Ref.~\cite{PDG2020}.

\begin{figure}
\begin{center}
\begin{minipage}[b]{0.32\textwidth}
\includegraphics[width=0.98\textwidth]{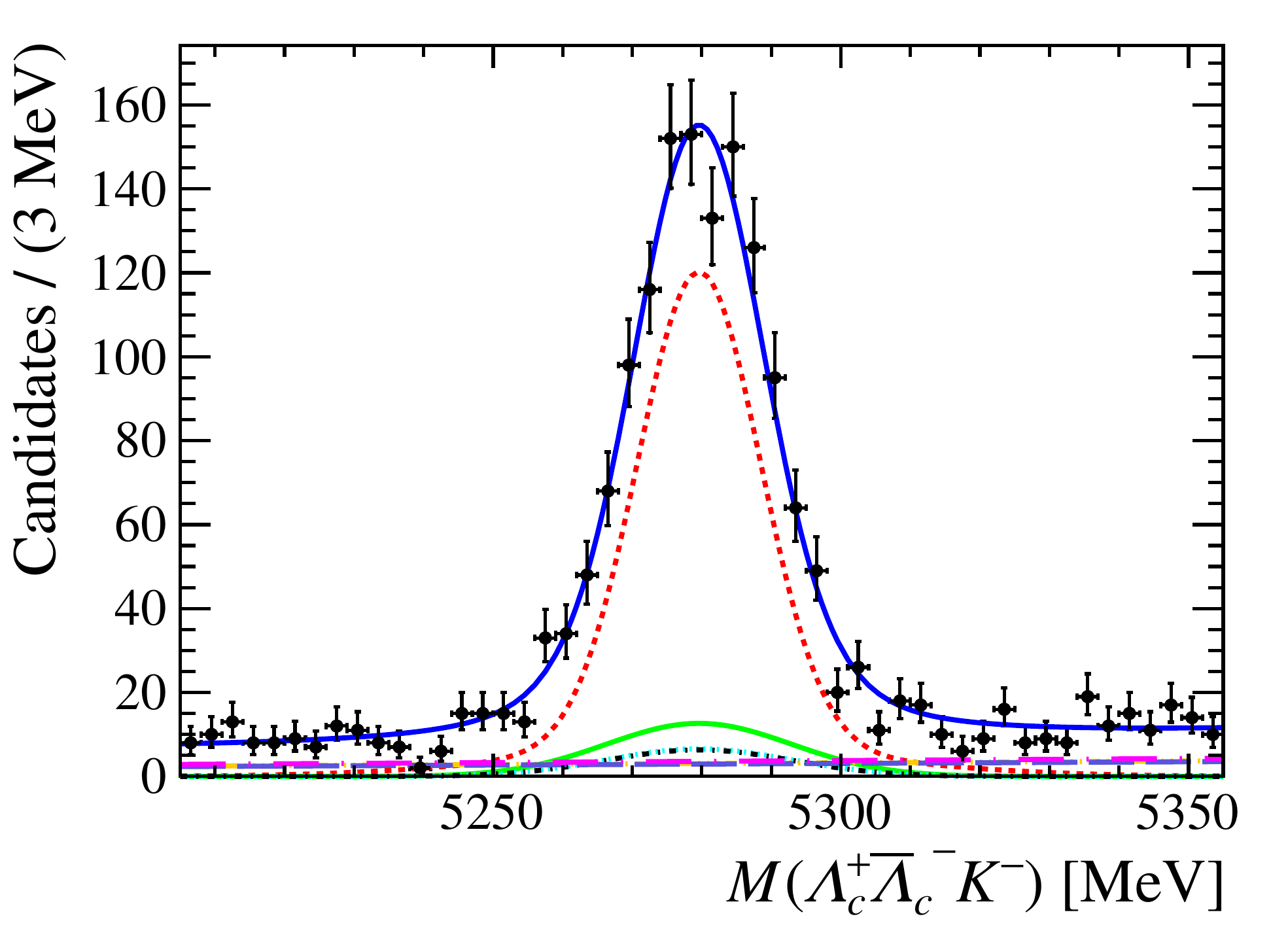}
\put(-115,80){LHCb}
\put(-115,65){5\,fb$^{-1}$}
\put(-30,80){(a)}
\end{minipage}
\begin{minipage}[b]{0.32\textwidth}
\includegraphics[width=0.98\textwidth]{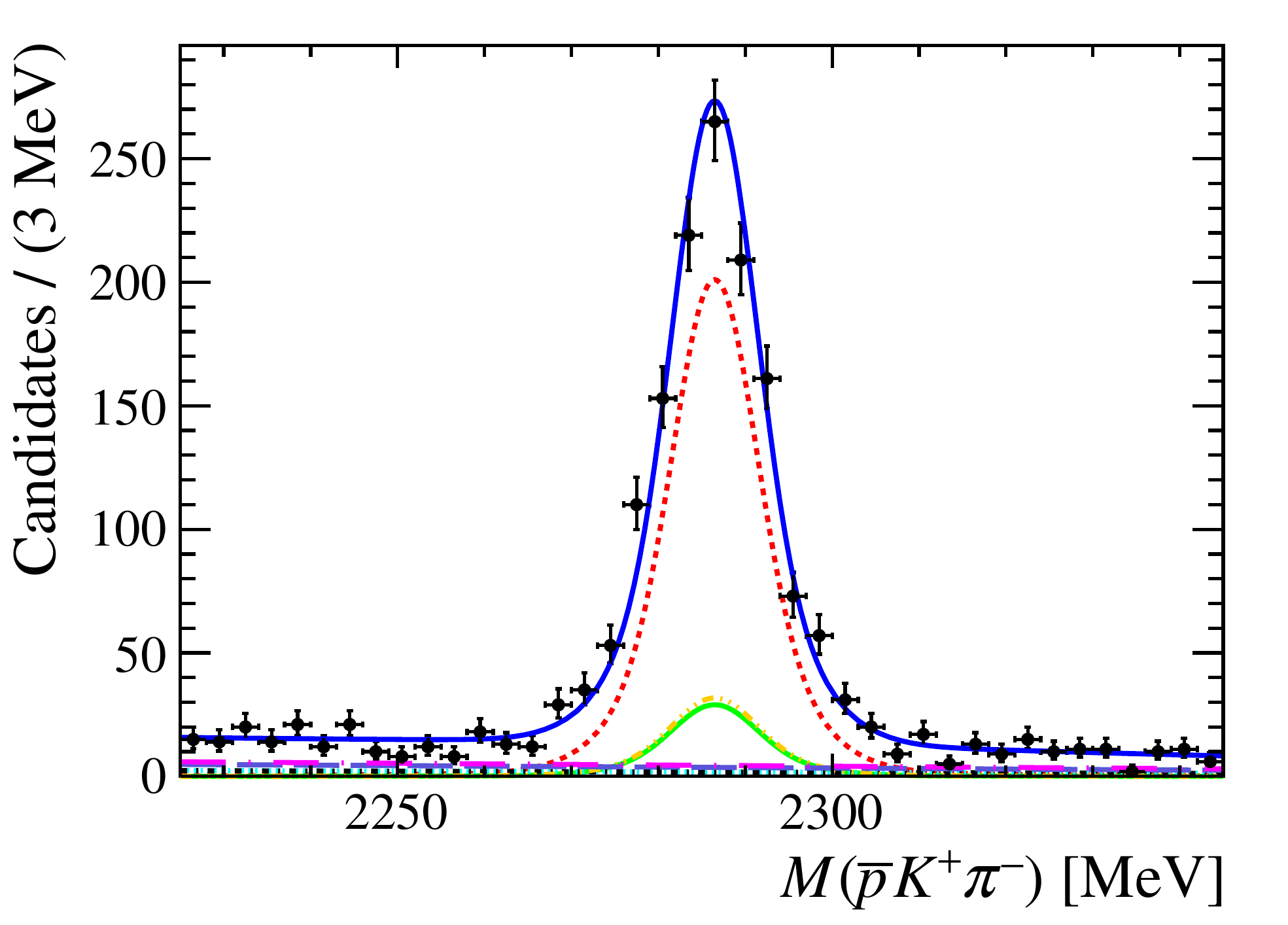}
\put(-115,80){LHCb}
\put(-115,65){5\,fb$^{-1}$}
\put(-30,80){(b)}
\end{minipage}
\begin{minipage}[b]{0.32\textwidth}
\includegraphics[width=0.98\textwidth]{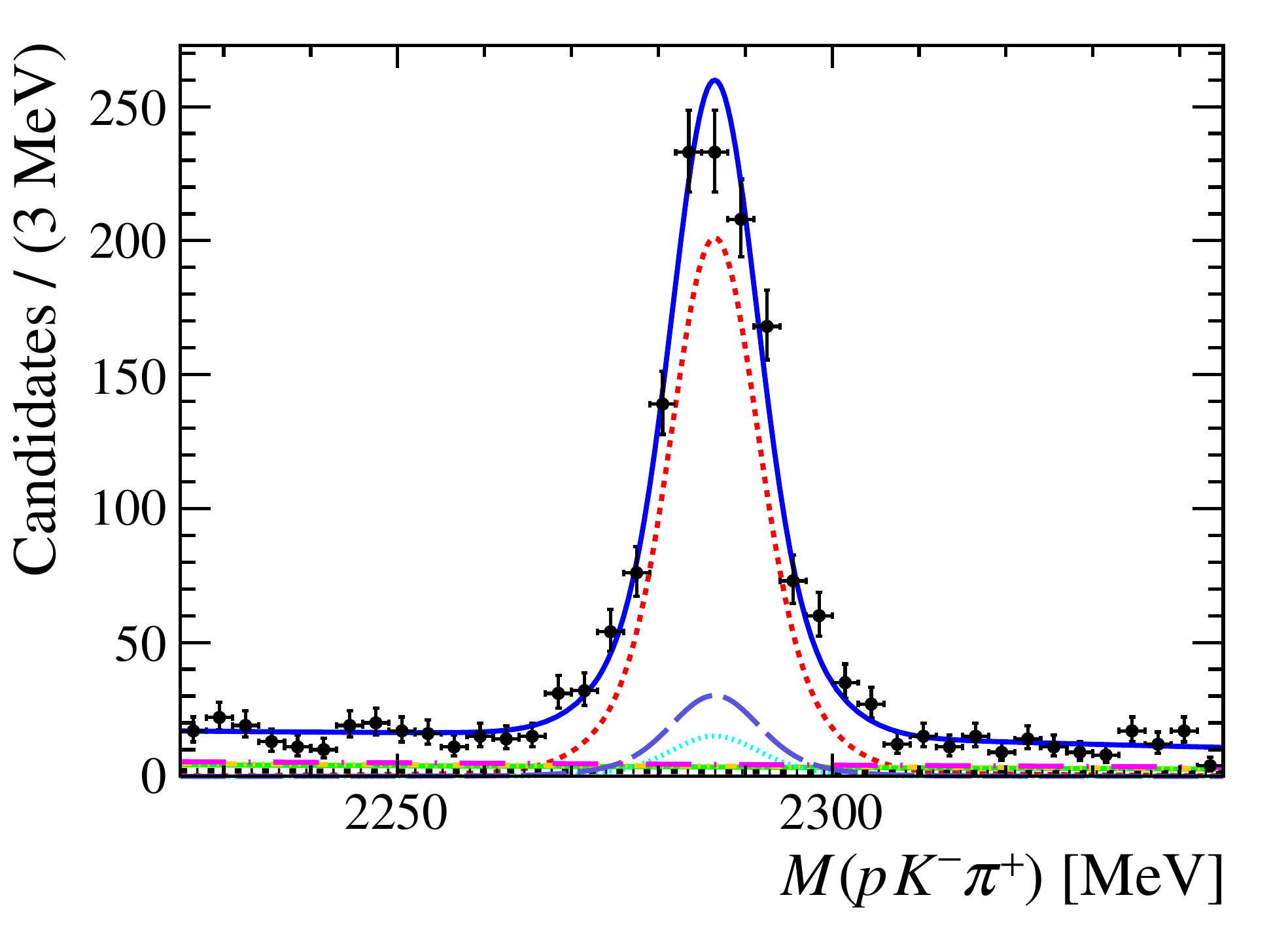}
\put(-115,80){LHCb}
\put(-115,65){5\,fb$^{-1}$}
\put(-30,80){(c)}
\end{minipage}
\begin{minipage}[b]{0.32\textwidth}
\includegraphics[width=0.98\textwidth]{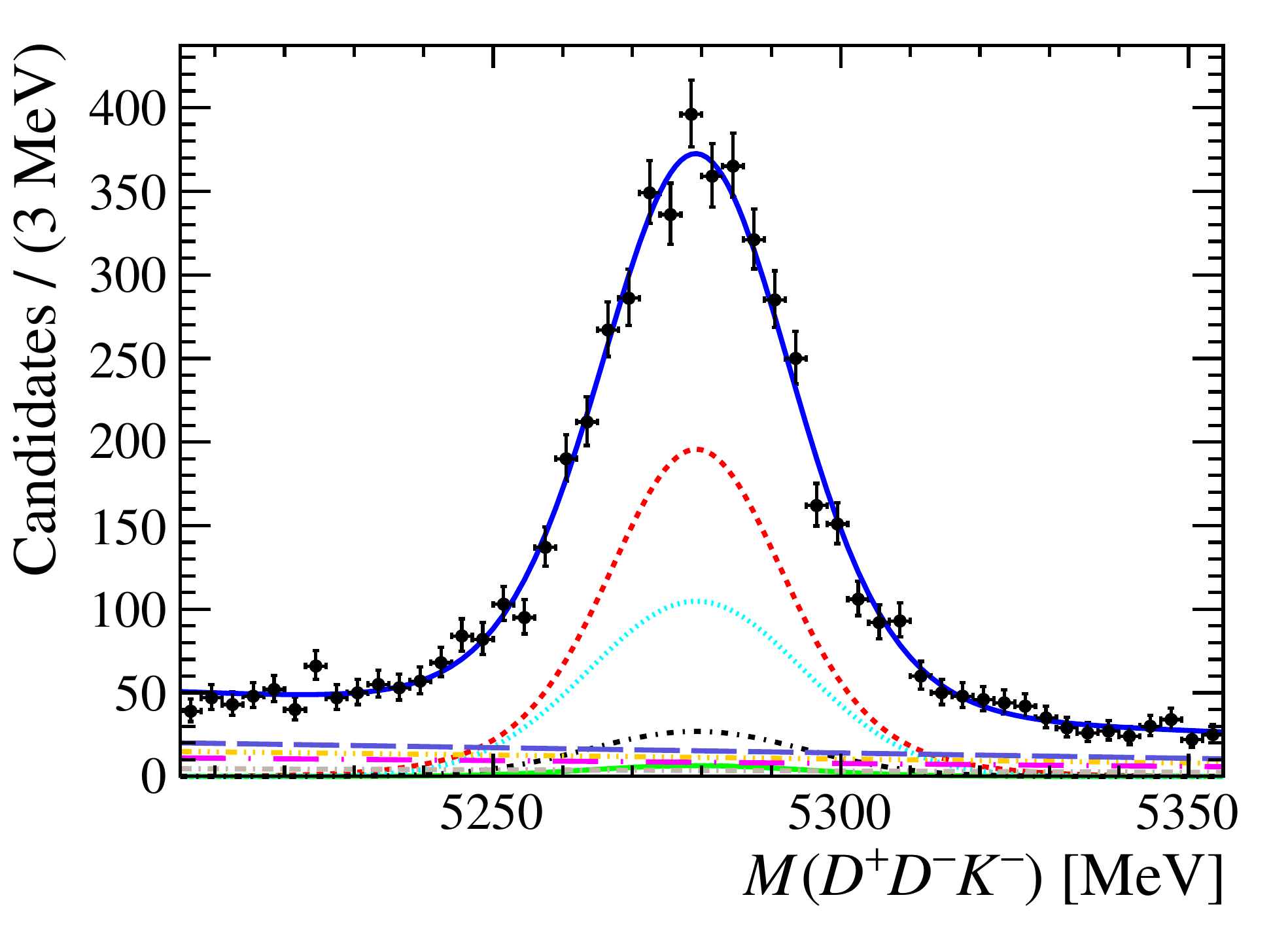}
\put(-115,80){LHCb}
\put(-115,65){5\,fb$^{-1}$}
\put(-30,80){(d)}
\end{minipage}
\begin{minipage}[b]{0.32\textwidth}
\includegraphics[width=0.98\textwidth]{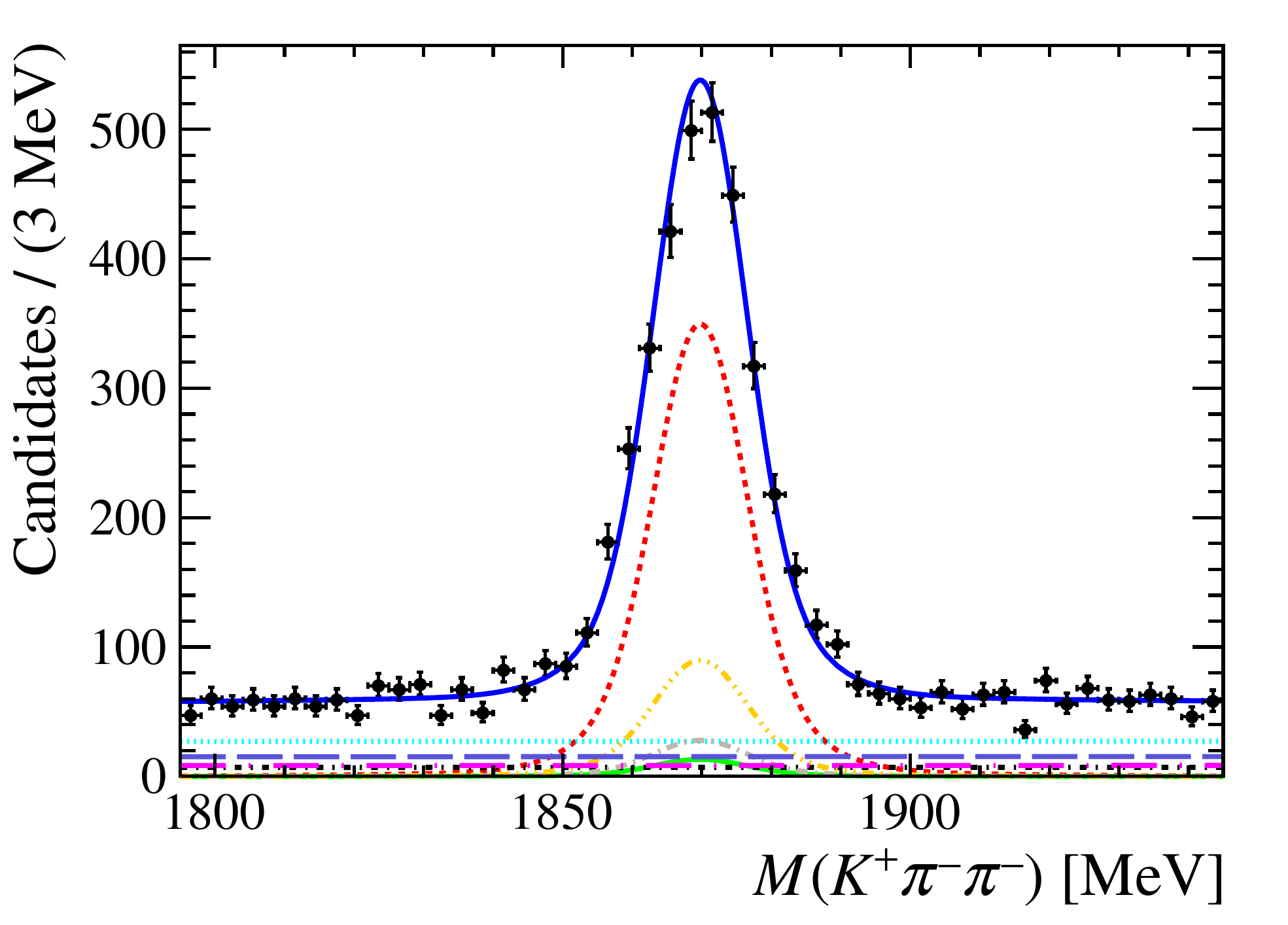}
\put(-115,80){LHCb}
\put(-115,65){5\,fb$^{-1}$}
\put(-30,80){(e)}
\end{minipage}
\begin{minipage}[b]{0.32\textwidth}
\includegraphics[width=0.98\textwidth]{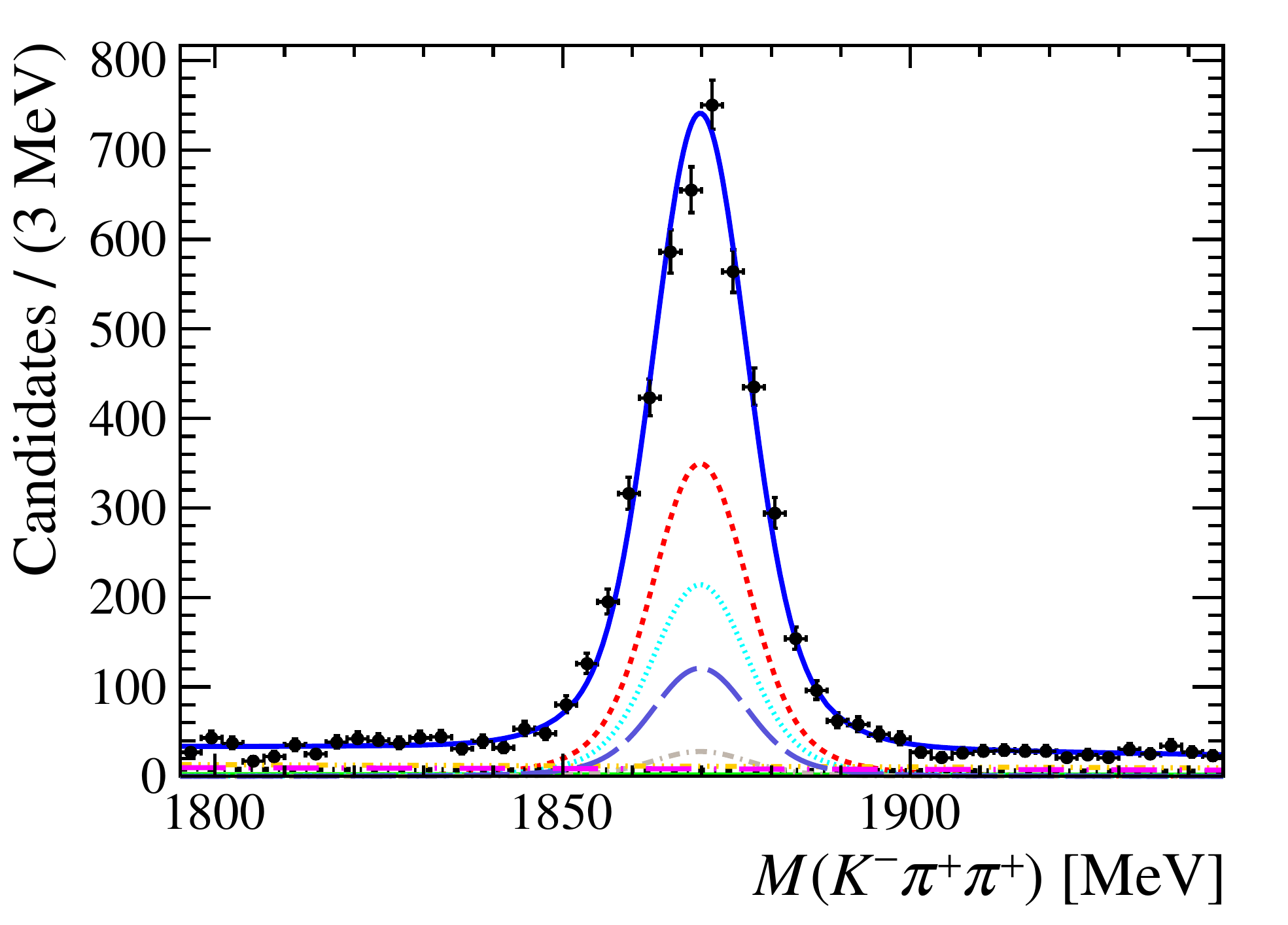}
\put(-115,80){LHCb}
\put(-115,65){5\,fb$^{-1}$}
\put(-30,80){(f)}
\end{minipage}
\caption{\small Distributions of (a) $M(\Lc\Lcbar\Km)$, (b) $M(\antiproton\Kp\pim)$ and (c) $M(\proton\Km\pip)$ in the  \mbox{$\BtoLcLcK$} sample and distributions of (d) $M(\Dm\Dp\Km)$, (e) $M(\Kp\pim\pim)$ and (f) $M(\Km\pip\pip)$ in the $\BtoDDK$ sample.
All selected events are triggered due to $\Bm$ candidates. The legend is the same as in Fig.~\ref{fig:Bu3Dfit}, except replacing $\Lc$ ($\Lcbar$) with $\Dp$ ($\Dm$) for the \mbox{$\BtoDDK$} decay.}
\label{fig:3Dfittrigger}
\end{center}
\end{figure}

In addition to the selection described in Sec.~\ref{sec:selection}, it is
required that all selected signal events are triggered on the $\Bm$ candidate decay products
for both signal and normalisation modes so that the trigger efficiencies
can be estimated accurately.
The projections of the 3D fit to the $\BtoLcLcK$ spectra with these additional requirements
are shown in Fig.~\ref{fig:3Dfittrigger} (a-c) and the measured yield is $977 \pm 36$.
For the normalisation decay $\BtoDDK$, a similar 3D fit is performed to 
the mass distributions of the $\Bm$, $\Dp$ and $\Dm$ candidates
with the same fitting functions except replacing $\Lz_{\cquark}$ with $D$
wherever applicable.
The fitted distributions are shown in Fig.~\ref{fig:3Dfittrigger} (d-f),
and the yield of $\BtoDDK$ decay is $2212 \pm 62$.

The total efficiency $\etot$ is the product of the efficiencies due to the detector
acceptance, trigger, reconstruction, offline preselection and multivariate selection.
The efficiencies from different sources are estimated with a combination of simulation and data.
Weights are assigned to the simulated events such that the track multiplicity distribution agrees with the data.
The tracking efficiency for charged tracks is estimated using data,
and the difference between simulation and data is obtained from a correction table
as a function of $\pt$, $\eta$ and number of charged tracks~\cite{Aaij:2014pwa}.
The total efficiencies of the signal and normalisation decays 
are $(3.41 \pm 0.02)\times 10^{-4}$ and $(8.15 \pm 0.11) \times 10^{-4}$,
respectively, where the uncertainties come from the statistics of the simulation samples.
An alternative set of selections of the $\BtoLcLcK$ signal 
with tighter criteria on $\pt$ and looser criteria on $\Bm$ decay time 
is used as a cross-check, and the efficiency-corrected yield is found to be consistent with the baseline selections.

The systematic uncertainty on the relative branching fraction comes from uncertainties
in signal yield determination and efficiency estimation.
The working point of the BDT output is chosen to maximise the signal significance.
The choice of BDT thresholds are varied and no significant bias is found.
The average of the efficiency-corrected yield agrees with the nominal result 
within statistical uncertainties, and the maximum change in the central value 
is taken as a systematic uncertainty.
Alternative assumptions are made when performing the 3D fits, including
varying the fitting range, using fixed shapes from simulation for the signal in the fit,
and varying the mass resolution for $B$ decay backgrounds.
The shifts in yield are taken as systematic uncertainties.
The simulation sample is weighted to match the track multiplicity distribution of data, and the uncertainties of the weights are propagated to the efficiencies.
The efficiency dependence on the decay phase space is studied in the $M(\Lcbar \Km)$ \vs
$M(\Lc\Km)$ ($M(\Dm\Km)$ \vs $M(\Dp\Km)$) plane for the signal (normalisation) decay.
The efficiencies are calculated alternatively according to
each candidate's decay phase space and the shift in central values is taken as a
systematic uncertainty.
The tracking efficiency has a negligible effect on the final results.
The PID values in simulation are sampled to better agree with data using
a dedicated tool~\cite{Aaij:2018vrk}. An alternative template is used for the sampling
and the shift in the efficiency-corrected yield is taken as a systematic uncertainty.
The parameterisation of the VELO materials, which influences IP-related variables, is tuned on simulation.
The $\chisqip$ distribution in simulation is scaled with parameters given by the $\Lb\to\Lc\pim$
sample, and the shift in $R_{\BF}$ using the scaled simulation is assigned as a
systematic uncertainty.
The trigger efficiencies are estimated using the simulation as the baseline. 
They are also estimated using data, taking the events triggered independent of the $\Bm$ candidates.
The estimated efficiencies between the two methods agree within statistical
uncertainties, and the difference between efficiencies calculated using data and
the nominal value is assigned as a systematic uncertainty.
The uncertainty due to the limited simulation sample size is propagated to the final result.
The systematic uncertainty on $R_{\BF}$ is summarised in Table~\ref{tab:sysbr},
and all the terms mentioned above are added in quadrature.

The ratio of branching fractions is measured to be
\begin{align*}
    R_{\BF} = 2.36 \pm 0.11 \pm 0.22 \pm 0.25,
\end{align*}
where the uncertainties are statistical, systematic and that due to uncertainties on the $\Lc$ and $\Dp$ branching fractions. 
This measurement is significantly improved with respect to $R_{\BF} = 2.23 \pm 0.78$ calculated using values from Ref.~\cite{PDG2020}.

\begin{table}[t]
\begin{center}
\caption{\small Relative systematic uncertainties of the branching fractions of signal, normalisation channel, and their ratio $R_{\BF}$ (in percent). Correlation between the two channels are considered.}
\begin{tabular}{lccc} \hline
  Source  &  $\BtoLcLcK$  & $\BtoDDK$ & $R_{\BF}$ \\
  \hline
  BDT working point      &  1.5  &  0.4  &  1.6\\
  Fit range              &  1.2  &  0.6  &  1.3\\
  Detector resolution    &  0.0   &  0.3  & 0.3 \\
  Signal model           &  0.0   &  1.6  & 1.6 \\
  Simulation weighting &  2.6  &  3.8  &  4.6\\
  Decay phase space      &  0.1  &  0.2  & 0.3 \\
  PID resample         &  0.6  &  0.9  &  1.0\\
  Detector parameterisation   &  3.3     &    2.5    & 0.8 \\
  Trigger                &  5.6  &  4.9  &  7.5\\
  Simulation sample size  &  0.9  &  1.2  &  1.5\\
  \hline
  Total                  &        &        &  9.4 \\
  \hline 
\end{tabular}
\label{tab:sysbr}
\end{center}
\end{table}

\section{Summary}
\label{sec:summary}
The $\BtoLcLcK$ decay is studied for the first time in $pp$ collisions using data
collected by the LHCb experiment, corresponding to an integrated luminosity of $5\invfb$.
In the $\Lc\Km$ invariant mass spectrum, two neutral excited charm baryon states,
$\Xic(2923)^{0}$ and $\Xic(2939)^{0}$, are observed. Their masses and widths are in agreement with those of the states
observed in a prompt $\Lc\Km$ measurement~\cite{LHCb-PAPER-2020-004}. These new measurements confirm
that the $\Xic(2930)^{0}$ state observed in $\BtoLcLcK$ at $B$-factories is resolved into
two separate states.
No resonance structure is found at higher masses.
Evidence for the $\Xic(2880)^{0}$ state is found with a significance of $3.8\,\sigma$.
In addition, evidence of a new $\Xic(2790)^{0}\to\Lc\Km$ decay is found
with a significance \mbox{of $3.7\,\sigma$}.

The relative branching fraction of $\BtoLcLcK$ is measured with respect to
$\BtoDDK$ as
$R_{\BF} = 2.36 \pm 0.11 \pm 0.22 \pm 0.25$,
where the uncertainties are statistical, systematic and that due to the charm decay branching fraction. This is the most precise measurement 
to date of this ratio.

%
\section*{Acknowledgements}
%
%
\noindent We express our gratitude to our colleagues in the CERN
accelerator departments for the excellent performance of the LHC. We
thank the technical and administrative staff at the LHCb
institutes.
We acknowledge support from CERN and from the national agencies:
CAPES, CNPq, FAPERJ and FINEP (Brazil); 
MOST and NSFC (China); 
CNRS/IN2P3 (France); 
BMBF, DFG and MPG (Germany); 
INFN (Italy); 
NWO (Netherlands); 
MNiSW and NCN (Poland); 
MEN/IFA (Romania); 
MICINN (Spain); 
SNSF and SER (Switzerland); 
NASU (Ukraine); 
STFC (United Kingdom); 
DOE NP and NSF (USA).
We acknowledge the computing resources that are provided by CERN, IN2P3
(France), KIT and DESY (Germany), INFN (Italy), SURF (Netherlands),
PIC (Spain), GridPP (United Kingdom), 
CSCS (Switzerland), IFIN-HH (Romania), CBPF (Brazil),
Polish WLCG  (Poland) and NERSC (USA).
We are indebted to the communities behind the multiple open-source
software packages on which we depend.
Individual groups or members have received support from
ARC and ARDC (Australia);
Minciencias (Colombia);
AvH Foundation (Germany);
EPLANET, Marie Sk\l{}odowska-Curie Actions and ERC (European Union);
A*MIDEX, ANR, IPhU and Labex P2IO, and R\'{e}gion Auvergne-Rh\^{o}ne-Alpes (France);
Key Research Program of Frontier Sciences of CAS, CAS PIFI, CAS CCEPP, 
Fundamental Research Funds for the Central Universities, 
and Sci. \& Tech. Program of Guangzhou (China);
GVA, XuntaGal, GENCAT and Prog.~Atracci\'on Talento, CM (Spain);
SRC (Sweden);
the Leverhulme Trust, the Royal Society
 and UKRI (United Kingdom).

%
%


\addcontentsline{toc}{section}{References}
\bibliographystyle{LHCb}
\bibliography{main,standard,LHCb-PAPER,LHCb-CONF,LHCb-DP,LHCb-TDR}

\newpage
\centerline
{\large\bf LHCb collaboration}
\begin
{flushleft}
\small
R.~Aaij$^{32}$\lhcborcid{0000-0003-0533-1952},
A.S.W.~Abdelmotteleb$^{50}$\lhcborcid{0000-0001-7905-0542},
C.~Abellan~Beteta$^{44}$,
F.~Abudin{\'e}n$^{50}$\lhcborcid{0000-0002-6737-3528},
T.~Ackernley$^{54}$\lhcborcid{0000-0002-5951-3498},
B.~Adeva$^{40}$\lhcborcid{0000-0001-9756-3712},
M.~Adinolfi$^{48}$\lhcborcid{0000-0002-1326-1264},
P.~Adlarson$^{77}$\lhcborcid{0000-0001-6280-3851},
H.~Afsharnia$^{9}$,
C.~Agapopoulou$^{13}$\lhcborcid{0000-0002-2368-0147}
% ...

{\footnotesize \it

% ...
$^{9}$Universit{\'e} Clermont Auvergne, CNRS/IN2P3, LPC, Clermont-Ferrand, France\\
% ...
$^{13}$LPNHE, Sorbonne Universit{\'e}, Paris Diderot Sorbonne Paris Cit{\'e}, CNRS/IN2P3, Paris, France\\
% ...
$^{32}$Nikhef National Institute for Subatomic Physics, Amsterdam, Netherlands\\
% ...
$^{40}$Instituto Galego de F{\'\i}sica de Altas Enerx{\'\i}as (IGFAE), Universidade de Santiago de Compostela, Santiago de Compostela, Spain\\
% ...
$^{44}$Physik-Institut, Universit{\"a}t Z{\"u}rich, Z{\"u}rich, Switzerland\\
% ...
$^{48}$H.H. Wills Physics Laboratory, University of Bristol, Bristol, United Kingdom\\
% ...
$^{50}$Department of Physics, University of Warwick, Coventry, United Kingdom\\
% ...
$^{53}$School of Physics and Astronomy, University of Glasgow, Glasgow, United Kingdom\\
$^{54}$Oliver Lodge Laboratory, University of Liverpool, Liverpool, United Kingdom\\
% ...
$^{77}$Department of Physics and Astronomy, Uppsala University, Uppsala, Sweden, associated to $^{53}$\\
% ...
}
\end{flushleft}

\end{document}